\documentclass[12pt]{article}
\usepackage{amsfonts,graphicx,amsmath,amsthm,amssymb,epsfig}
\usepackage{float}
\allowdisplaybreaks
\begin{document}

\title{\bf Charged Anisotropic Spherical Collapse in $f(\mathcal{R},\mathcal{T},\mathcal{Q})$ Gravity}
\author{M. Sharif$^1$ \thanks{msharif.math@pu.edu.pk}~ and
Tayyab Naseer$^{1,2}$ \thanks{tayyabnaseer48@yahoo.com; tayyab.naseer@math.uol.edu.pk}\\
$^1$ Department of Mathematics and Statistics, The University of Lahore,\\
1-KM Defence Road Lahore, Pakistan.\\
$^2$ Department of Mathematics, University of the Punjab,\\
Quaid-i-Azam Campus, Lahore-54590, Pakistan.}

\date{}
\maketitle

\begin{abstract}
This paper discusses the gravitational collapse of dynamical
self-gravitating fluid distribution in
$f(\mathcal{R},\mathcal{T},\mathcal{Q})$ gravity, where
$\mathcal{Q}=\mathcal{R}_{\varphi\vartheta}\mathcal{T}^{\varphi\vartheta}$.
In this regard, we assume a charged anisotropic spherical geometry
involving dissipation flux and adopt standard model of the form
$\mathcal{R}+\Phi\sqrt{\mathcal{T}}+\Psi\mathcal{Q}$, where $\Phi$
and $\Psi$ symbolize real-valued coupling parameters. The
Misner-Sharp as well as M\"{u}ler-Israel Stewart mechanisms are
employed to formulate the corresponding dynamical and transport
equations. We then interlink these evolution equations which help to
study the impact of state variables, heat dissipation, modified
corrections and charge on the collapse rate. The Weyl scalar is
further expressed in terms of the modified field equations. The
necessary and sufficient condition of conformal flatness of the
considered configuration and homogeneous energy density is obtained
by applying some constraints on the model along with disappearing
charge and anisotropy. Finally, we discuss different cases to
investigate how the spherical matter source is affected by the
charge and modified corrections.
\end{abstract}

\section{Introduction}

Cosmological observations indicate that the luminous plasma in our
universe (held together by its own gravity and known as a star)
originated when the superheated matter and energy expanded. Various
astronomers have played their ample attention in order to study the
evolutionary phases of such objects. After the big bang, the lumps
of masses, primarily comprising of helium and hydrogen, began to
contract due to the attractive nature of gravity. This contraction
produces pressure and heat due to which hydrogen atoms started to
fuse. A star releases a heavy amount of energy due to the fusion
process. As a result, an outward pressure is generated that helps to
neutralize the strong gravitational pull towards the center of that
star. Since limited fuel (in the form of hydrogen atoms) is
available, therefore, the equilibrium of the two forces (i.e.,
inward gravity and outward pressure) does not last forever.
Consequently, a star shrinks its radius and reaches its dense state
leading to a collapsing phenomenon. It is a fundamental, captivating
and promising phenomenon for the formation of new dense objects,
like white dwarfs, neutron stars and black holes. The geometrical
structures of galaxies can be different, however, the spherical
geometry has the most impressive characteristics due to its
transition to other stellar structures.

The importance of this phenomenon in the field of relativistic
astrophysics was disclosed by the pioneering work of Chandrasekhar
\cite{28}. According to him, the counterbalance of the attractive
force of a compact star due to its mass and outward directed
pressure (produces as a result of nuclear fusions) keeps that body
stable. Oppenheimer and Snyder \cite{29} explored the collapse of
dynamical dust sphere and observed the ultimate formation of a black
hole. Misner and Sharp \cite{29a} analyzed the role of anisotropic
pressure on the collapse rate of spherical geometry and found the
corresponding mass density transforming into the heat flux. Later,
the relativistic models suggested by Glass \cite{ak} were studied by
Santos \cite{aj}, in which he calculated the junction conditions for
a spherical shearfree interior fluid distribution. The Misner-Sharp
approach was used by Herrera and Santos \cite{30} to discuss
spherical dissipative collapse whose energy dissipates by means of
heat and radiations.

The evolutionary pattern and stability of massive objects can
significantly be studied in the presence of electromagnetic field.
The influence of such a field produces magnetic and Coulomb
(repulsive) forces that help to reduce the gravitational attraction
of an object. Thus, a star requires more charge to be in stable form
in a strong gravitational field. The impact of electric charge on
the dynamical instability of isotropic configuration has been
studied by Glazer \cite{ac}. Zhang \emph{et al.} \cite{ad} showed
that the structure of a neutron star may change for the case when
the interior charge is equivalent to the mass density. Bekenstein
\cite{31a} analyzed the collapse of charged spherical spacetime and
observed a notable reduction in the collapse rate. The same results
have been found by Esculpi and Aloma \cite{31b} while studying
anisotropic sphere influenced from electromagnetic field. Pinheiro
and Chan \cite{31c} concluded from the collapse of spherical
self-gravitating object that the presence of charge diminishes the
probability of formation of the Reissner-Nordstr\"{o}m black hole.
Sharif and his collaborators \cite{32}-\cite{32c} studied different
self-gravitating configurations with/without including the effect of
charge and deduced that the collapse rate of charged bodies is much
lesser than that of uncharged systems. Another important factor in
the evolution of stellar structure that is used to measure its
geometric curvature is known as the Weyl tensor. Penrose \cite{36a}
interlinked various physical variables such as the pressure and
energy density with the Weyl tensor and employed these relations to
study spherical collapse. In this scenario, the relations between
pressure anisotropy, the Weyl tensor and energy density have also
been established corresponding to spherical and cylindrical
geometries \cite{36,36aaa}. Sharif and Fatima \cite{36b} studied
cylindrically symmetric spacetime by exploring such relation and
obtained if and only if condition between conformal flatness and the
energy density homogeneity.

A factor that exclusively affects the gravitational collapse is the
dissipation, thus its effects cannot be ruled out in the study of
collapsing phenomenon \cite{32d,33}. Chan \cite{34,34b} investigated
a radiating geometry coupled with isotropic distribution and
revealed that there appears more and more anisotropy due to the
presence of shear viscosity. The collapsing time of the geometries
with and without shear viscosity is found to be the same. Di Prisco
\emph{et al.} \cite{35c} analyzed anisotropic sphere in the presence
of shear and charge, and disclosed the impact of these quantities on
the energy density inhomogeneity. The collapse rate of
quasi-spherical Szekeres interior along with charged Vaidya exterior
spacetime has been discussed by Nath \emph{et al.} \cite{35a}
through matching conditions. They also deduced that the presence of
charge supports the formation of naked singularity. Herrera \emph{et
al.} \cite{35} studied viscous dissipative matter distribution and
observed that the attractive force of gravity is lessened by the
governing parameters, and thus the collapse rate diminishes. Sharif
and Abbas \cite{c} explored a five-dimensional spherical collapse in
the presence of electromagnetic field and the cosmological constant.
Sharif and Bhatti \cite{d} coupled dynamical and transport equations
to check the effects of dissipation and electromagnetic field over
collapsing process. The collapse of shearfree anisotropic spherical
models has also been discussed in \cite{f}.

The ongoing accelerated expansion of our cosmos have become an
enticing topic in the field of astrophysics and cosmology for the
last few decades \cite{1a,1b}. Multiple observations have been done
and it was found that a mysterious force with immense repulsive
nature, namely dark energy, triggers such expansion as it comprises
more than $70\%$ of the whole universe. The most favorable theory to
discuss comic structures and their evolution was considered as the
general theory of relativity ($\mathbb{GR}$), however, astronomers
faced two major short comings in this theory while studying cosmic
expansion such as the fine-tuning and coincidence problem. Later,
many modifications of $\mathbb{GR}$ have been introduced as a
suitable solution to these problems. The first ever extension (also
considered as the straightforward generalization) of $\mathbb{GR}$
to disclose cosmic features at large scales, is the $f(\mathcal{R})$
theory that was obtained by inserting generic functional of the
Ricci scalar in place of $\mathcal{R}$ in an Einstein-Hilbert action
\cite{1c}. The physically acceptable outcomes corresponding to
various $f(\mathcal{R})$ models have been obtained in the literature
\cite{2}-\cite{2d}.

The appealing nature and composition of this universe can be well
understood by coupling the matter distribution and geometry.
Bertolami \emph{et al.} \cite{10} initially presented this idea in
$f(\mathcal{R})$ framework by engaging the Ricci scalar in the
matter Lagrangian and studied its effects on celestial bodies. The
idea of such interaction was recently generalized at the action
level by Harko \emph{et al.} \cite{20} through pioneering
$f(\mathcal{R},\mathcal{T})$ theory, in which $\mathcal{T}$ is trace
of the energy-momentum tensor $(\mathbb{EMT})$. The effective
$\mathbb{EMT}$ of any extended theory whose functional involves the
entity $\mathcal{T}$, yields its non-vanishing divergence, in
contrast to $\mathbb{GR}$ and $f(\mathcal{R})$ gravity. Several
astrophysicists have employed different approaches in this scenario
and obtained remarkable results \cite{21d}-\cite{21f}. The
limitation of $f(\mathcal{R},\mathcal{T})$ gravity is that it does
not able to study the coupling effects on test particles under some
particular situations. Soon after this, Haghani \emph{et al.}
\cite{22} developed $f(\mathcal{R},\mathcal{T},\mathcal{Q})$ to
overcome this flaw. They also considered different matter
distributions (such as pressureless fluid and the high density
regime) and discussed their cosmological applications by choosing
three different modified models.

Sharif and Zubair \cite{22a} chose the matter Lagrangian in terms of
pressure and energy density, and studied first two black hole
thermodynamics laws corresponding to two models such as
$\mathcal{R}+\lambda\mathcal{Q}$ and
$\mathcal{R}(1+\lambda\mathcal{Q})$. The energy bounds for the same
models have also been addressed by them from which different
acceptable values of the coupling constant $\lambda$ are deduced
\cite{22b}. Odintsov and S\'{a}ez-G\'{o}mez \cite{23} assumed the
flat FLRW metric to discuss cosmological implications for several
$f(\mathcal{R},\mathcal{T},\mathcal{Q})$ models and the
corresponding issue of fluid instability. They reconstructed the
action for this theory and also presented de Sitter solutions.
Baffou \emph{et al.} \cite{25} discussed the above models and
observed their stability from the convergence of geometrical
perturbation functions and matter distribution. We have recently
discussed multiple anisotropic spherical structures in the absence
and presence of electromagnetic field through various approaches and
found physically acceptable solutions \cite{27}-\cite{27a5}. The
structure scalars have also been deduced in modified scenario to
discuss the evolution of different static/non-static geometries
\cite{27aa}-\cite{27aaa3}.

The phenomenon of the gravitational collapse of spherical spacetime
with different matter distributions has also been explored in the
context of multiple extended theories. Borisov \emph{et al.}
\cite{27ab} studied spherical collapse in $f(\mathcal{R})$ gravity
by employing numerical simulations. Chakrabarti and Banerjee
\cite{27ac} explored spherical collapse in the same theory and
concluded that even a vacuum collapse may result in the formation of
a black hole. The dynamical equations have been extensively used in
the study of collapsing rate corresponding to anisotropic
self-gravitating matter sources in the background of
$f(\mathcal{R},\mathcal{T})$ theory \cite{27ad}-\cite{27ad2}. It was
concluded that additional terms of this modified gravity enhances
stability of the considered distribution. Bhatti \emph{et al.}
\cite{27aad,27aad1} investigated spherical anisotropic geometry
coupled with dissipation flux and discussed the role of different
forces on the collapse rate in
$f(\mathcal{R},\mathcal{T},\mathcal{Q})$ framework. This concept has
been studied for different matter sources with and without involving
charge in several extended theories \cite{27af}-\cite{27af2}. They
observed that the charged system becomes more stable as the rate of
collapse reduces in the presence of electromagnetic field.

This article is devoted to the study of gravitational collapse of
the dynamical anisotropic dissipative sphere in the presence of an
electromagnetic field in the background of
$f(\mathcal{R},\mathcal{T},\mathcal{R}_{\phi\psi}\mathcal{T}^{\phi\psi})$
theory. The structure of this paper is given as follows. Some
fundamental quantities and the modified field equations analogous to
the model $\mathcal{R}+\Phi\sqrt{\mathcal{T}}+\Psi\mathcal{Q}$ are
formulated in the next section. The two non-zero components of the
Bianchi identities and their connection with the acceleration of
considered source are established in Section \textbf{3}. We then
develop the transport equation and identify some dynamical forces to
analyze the increment/decrement in the collapse rate in Section
\textbf{4}. Section \textbf{5} presents the relationship among the
Weyl scalar, effective physical quantities and charge. Section
\textbf{6} concludes all our results.

\section{The $f(\mathcal{R},\mathcal{T},\mathcal{R}_{\varphi\vartheta}\mathcal{T}^{\varphi\vartheta})$ Gravity}

The insertion of modified functional
$f(\mathcal{R},\mathcal{T},\mathcal{Q})$ in an Einstein-Hilbert
action (with $\kappa=8\pi$) leads to \cite{23}
\begin{equation}\label{g1}
\mathbb{S}_{f(\mathcal{R},\mathcal{T},\mathcal{Q})}=\int\sqrt{-g}
\left\{\frac{f(\mathcal{R},\mathcal{T},\mathcal{Q})}{16\pi}+\mathbb{L}_{\mathcal{E}}
+\mathbb{L}_{\mathfrak{M}}\right\}d^{4}x,
\end{equation}
where the Lagrangian densities for the electromagnetic field and the
fluid configuration are represented by $\mathbb{L}_{\mathcal{E}}$
and $\mathbb{L}_{\mathfrak{M}}$, respectively. Implementation of the
variational principle on the action \eqref{g1} produces the field
equations as
\begin{equation}\label{g2}
\mathcal{G}_{\varphi\vartheta}=\mathcal{T}_{\varphi\vartheta}^{(\mathrm{EFF})}=\frac{1}
{f_{\mathcal{R}}-\mathbb{L}_{\mathfrak{M}}f_{\mathcal{Q}}}\left\{8\pi\big(\mathcal{T}_{\varphi\vartheta}+\mathcal{E}_{\varphi\vartheta}\big)
+\mathcal{T}_{\varphi\vartheta}^{(D)}\right\}.
\end{equation}
Here, $\mathcal{G}_{\varphi\vartheta}$ and
$\mathcal{T}_{\varphi\vartheta}^{(\mathrm{EFF})}$ indicate the
Einstein tensor and the $\mathbb{EMT}$ corresponding to extended
theory, respectively. Also, $\mathcal{T}_{\varphi\vartheta}$ is the
anisotropic fluid distribution in this case and
$\mathcal{E}_{\varphi\vartheta}$ is the electromagnetic field
tensor. The modification of gravity triggers the factor
$\mathcal{T}_{\varphi\vartheta}^{(D)}$ which has the following form
\begin{eqnarray}\nonumber
\mathcal{T}_{\varphi\vartheta}^{(D)}&=&\left(f_{\mathcal{T}}+\frac{1}{2}\mathcal{R}f_{\mathcal{Q}}\right)\mathcal{T}_{\varphi\vartheta}
+\left\{\frac{\mathcal{R}}{2}\left(\frac{f}{\mathcal{R}}-f_{\mathcal{R}}\right)
-\frac{1}{2}\nabla_{\varrho}\nabla_{\omega}\big(f_{\mathcal{Q}}\mathcal{T}^{\varrho\omega}\big)\right.\\\nonumber
&-&\left.\mathbb{L}_{\mathfrak{M}}f_{\mathcal{T}}\right\}g_{\varphi\vartheta}
-\frac{1}{2}\Box\big(f_{\mathcal{Q}}\mathcal{T}_{\varphi\vartheta}\big)-2f_{\mathcal{Q}}\mathcal{R}_{\varrho(\varphi}
\mathcal{T}_{\vartheta)}^{\varrho}+\nabla_{\varrho}\nabla_{(\varphi}[\mathcal{T}_{\vartheta)}^{\varrho}
f_{\mathcal{Q}}]\\\label{g4}
&-&\big(g_{\varphi\vartheta}\Box-\nabla_{\varphi}\nabla_{\vartheta}\big)f_{\mathcal{R}}+2\big(f_{\mathcal{Q}}\mathcal{R}^{\varrho\omega}
+f_{\mathcal{T}}g^{\varrho\omega}\big)\frac{\partial^2\mathbb{L}_{\mathfrak{M}}}{\partial
g^{\varphi\vartheta}\partial g^{\varrho\omega}},
\end{eqnarray}
where $f_{\mathcal{R}}=\frac{\partial
f(\mathcal{R},\mathcal{T},\mathcal{Q})}{\partial
\mathcal{R}}$,~$f_{\mathcal{T}}=\frac{\partial
f(\mathcal{R},\mathcal{T},\mathcal{Q})}{\partial \mathcal{T}}$ and
$f_{\mathcal{Q}}=\frac{\partial
f(\mathcal{R},\mathcal{T},\mathcal{Q})}{\partial \mathcal{Q}}$.
Also, $\nabla_\varrho$ and $\Box\equiv
\frac{1}{\sqrt{-g}}\partial_\varphi\big(\sqrt{-g}g^{\varphi\vartheta}\partial_{\vartheta}\big)$
are identified as the covariant derivative and the D'Alembert
operator, respectively. The matter Lagrangian is generally observed
to be in terms of pressure or energy density of the fluid in the
literature. However, the presence of charge in this case prompt us
to take its most suitable choice as
$\mathbb{L}_{\mathfrak{M}}=-\frac{1}{4}\mathcal{H}_{\varphi\vartheta}\mathcal{H}^{\varphi\vartheta}$
leading to $\frac{\partial^2\mathbb{L}_{\mathfrak{M}}} {\partial
g^{\varphi\vartheta}\partial
g^{\varrho\omega}}=-\frac{1}{2}\mathcal{H}_{\varphi\varrho}\mathcal{H}_{\vartheta\omega}$
\cite{22}. Here,
$\mathcal{H}_{\varphi\vartheta}=\varpi_{\vartheta;\varphi}-\varpi_{\varphi;\vartheta}$
is the Maxwell field tensor and
$\varpi_{\vartheta}=\varpi(r)\delta_{\vartheta}^{0}$ is the
four-potential.

To examine the collapse of dynamical spherical system, we take the
inner geometry represented by the metric
\begin{equation}\label{g6}
ds^2=-\mathcal{P}^2dt^2+\mathcal{W}^2dr^2+\mathcal{Y}^2\big(d\theta^2+\sin^2\theta
d\phi^2\big),
\end{equation}
where $\mathcal{P}=\mathcal{P}(t,r)$ and
$\mathcal{W}=\mathcal{W}(t,r)$ are dimensionless, and
$\mathcal{Y}=\mathcal{Y}(t,r)$ has the dimension of length. The
considered geometry is coupled with anisotropic fluid as well as
heat dissipation, whose $\mathbb{EMT}$ is given as
\begin{align}\nonumber
\mathcal{T}_{\varphi\vartheta}&=\big(\mu+\mathrm{P}_{\bot}\big)\mathcal{U}_{\varphi}\mathcal{U}_{\vartheta}
-\mathrm{P}_{\bot}g_{\varphi\vartheta}-\big(\mathrm{P}_{\bot}-\mathrm{P}_{\mathrm{r}}\big)
\mathcal{K}_{\varphi}\mathcal{K}_{\vartheta}\\\label{g5}
&+\varsigma_{\varphi}\mathcal{U}_\vartheta+\varsigma_{\vartheta}\mathcal{U}_\varphi-\big(g_{\varphi\vartheta}
+\mathcal{U}_{\varphi}\mathcal{U}_{\vartheta}\big)\zeta\Theta,
\end{align}
where $\Theta$ is the expansion scalar and and $\zeta$ denotes the
coefficient of bulk viscosity. The state variables such as the
energy density, radial and tangential pressures are symbolized by
$\mu,~\mathrm{P}_{\mathrm{r}}$ and $\mathrm{P}_{\bot}$,
respectively. Other quantities like heat flux
($\varsigma_{\varphi}$), the four-velocity
($\mathcal{U}_{\varphi}$), four-vector ($\mathcal{K}_{\varphi}$) and
$\Theta$ are demonstrated as
\begin{align}
\mathcal{U}_{\varphi}=-\mathcal{P}\delta_{\varphi}^{0},\quad
\mathcal{K}_{\varphi}=\mathcal{W}\delta_{\varphi}^{1},\quad
\Theta=\mathcal{U}^{\varphi}_{;\varphi},\quad
\varsigma_{\varphi}=\varsigma\mathcal{B}\delta_{\varphi}^{1},
\end{align}
satisfying the relations
\begin{align}
\mathcal{U}_{\varphi}\mathcal{U}^{\varphi}=-1,\quad
\mathcal{K}_{\varphi}\mathcal{K}^{\varphi}=1,\quad
\mathcal{U}_{\varphi} \mathcal{K}^{\varphi}=0.
\end{align}
The divergence of the $\mathbb{EMT}$ analogous to this extended
theory is observed to be non-zero, i.e., $\nabla_\varphi
\mathcal{T}^{\varphi\vartheta}\neq 0$, due to the presence of
matter-geometry coupling. Such coupling results in the origination
of an extra force which alters the motion of test particles from
geodesic path in their gravitational field. Thus we obtain
\begin{align}\nonumber
\nabla^\varphi\big(\mathcal{T}_{\varphi\vartheta}+\mathcal{E}_{\varphi\vartheta}\big)&=\frac{2}{2f_\mathcal{T}+\mathcal{R}f_\mathcal{Q}
+16\pi}\bigg[\nabla_\varphi\big(f_\mathcal{Q}\mathcal{R}^{\varrho\varphi}\mathcal{T}_{\varrho\vartheta}\big)
+\nabla_\vartheta\big(\mathbb{L}_\mathfrak{M}f_\mathcal{T}\big)\\\nonumber
&-\mathcal{G}_{\varphi\vartheta}\nabla^\varphi\big(f_\mathcal{Q}\mathbb{L}_\mathfrak{M}\big)-\frac{1}{2}\nabla_\vartheta
\mathcal{T}^{\varrho\omega}\big(f_\mathcal{T}g_{\varrho\omega}+f_\mathcal{Q}\mathcal{R}_{\varrho\omega}\big)\\\label{g4a}
&-\frac{1}{2}\big\{\nabla^{\varphi}(\mathcal{R}f_{\mathcal{Q}})+2\nabla^{\varphi}f_{\mathcal{T}}\big\}\mathcal{T}_{\varphi\vartheta}\bigg].
\end{align}
The trace of $f(\mathcal{R},\mathcal{T},\mathcal{Q})$ field
equations has the following form
\begin{align}\nonumber
&3\nabla^{\varrho}\nabla_{\varrho}
f_\mathcal{R}-\mathcal{R}\left(\frac{\mathcal{T}}{2}f_\mathcal{Q}-f_\mathcal{R}\right)-\mathcal{T}(8\pi+f_\mathcal{T})+\frac{1}{2}
\nabla^{\varrho}\nabla_{\varrho}(f_\mathcal{Q}\mathcal{T})\\\nonumber
&+\nabla_\varphi\nabla_\varrho(f_\mathcal{Q}\mathcal{T}^{\varphi\varrho})-2f+(\mathcal{R}f_\mathcal{Q}+4f_\mathcal{T})\mathbb{L}_\mathfrak{M}
+2\mathcal{R}_{\varphi\varrho}\mathcal{T}^{\varphi\varrho}f_\mathcal{Q}\\\nonumber
&-2g^{\vartheta\xi}
\frac{\partial^2\mathbb{L}_\mathfrak{M}}{\partial
g^{\vartheta\xi}\partial
g^{\varphi\varrho}}\left(f_\mathcal{T}g^{\varphi\varrho}+f_\mathcal{Q}\mathcal{R}^{\varphi\varrho}\right)=0.
\end{align}
For $f_\mathcal{Q}=0$, the field equations reduce to
$f(\mathcal{R},\mathcal{T})$ theory which further reduces to
$f(\mathcal{R})$ for the vacuum case.

The $\mathbb{EMT}$ for electromagnetic field is defined as
\begin{equation*}
\mathcal{E}_{\varphi\vartheta}=\frac{1}{4\pi}\left[\frac{1}{4}g_{\varphi\vartheta}\mathcal{H}^{\varrho\xi}\mathcal{H}_{\varrho\xi}
-\mathcal{H}^{\xi}_{\varphi}\mathcal{H}_{\xi\vartheta}\right].
\end{equation*}
The Maxwell equations are given as
\begin{equation*}
\mathcal{H}^{\varphi\vartheta}_{;\vartheta}=4\pi
\mathcal{J}^{\varphi}, \quad
\mathcal{H}_{[\varphi\vartheta;\varrho]}=0,
\end{equation*}
where $\mathcal{J}^{\varphi}=\rho \mathcal{K}^{\varphi}$, $\rho$ and
$\mathcal{J}^{\varphi}$ define the charge and current densities,
respectively. The above (left) equation takes the following
differential form
\begin{equation*}
\varpi''-\bigg(\frac{\mathcal{P}'}{\mathcal{P}}+\frac{\mathcal{W}'}{\mathcal{W}}
-\frac{2\mathcal{Y}'}{\mathcal{Y}}\bigg)\varpi'=4\pi\rho\mathcal{PW}^2,
\end{equation*}
providing
\begin{equation}\nonumber
\varpi'=\frac{s\mathcal{PW}}{\mathcal{Y}^2}.
\end{equation}
The expression $s=4\pi\int\rho\mathcal{WY}^2dr$ ensures the presence
of charge in the interior of self-gravitating distribution. Also,
$'=\frac{\partial}{\partial r}$. We thus get the matter Lagrangian
as $\mathbb{L}_{\mathfrak{M}}=\frac{s^2}{2\mathcal{Y}^4}$.

As $f(\mathcal{R},\mathcal{T},\mathcal{Q})$ theory produces much
complications in equations of motion, a linear model in this regard
is helpful in studying the current spherical source. The model is
given by
\begin{equation}\label{g5d}
f(\mathcal{R},\mathcal{T},\mathcal{Q})=f_1(\mathcal{R})+f_2(\mathcal{T})+f_3(\mathcal{Q})=\mathcal{R}+\Phi\sqrt{\mathcal{T}}+\Psi\mathcal{Q}.
\end{equation}
The values of coupling parameters that lie in their observed ranges
guarantee the physical acceptability of resulting solutions. The
model \eqref{g5d} (with $\Phi=0$) has been employed to examine some
isotropic configurations and their feasibility is observed for
particular values of $\Psi$ \cite{22a,22b}. The entities
$\mathcal{R},~\mathcal{T}$ and $\mathcal{Q}$ involving in this model
are
\begin{align}\nonumber
\mathcal{R}&=-\frac{2}{\mathcal{P}^3\mathcal{W}^3\mathcal{Y}^2}\bigg[2\mathcal{P}^3\mathcal{W}\mathcal{Y}\mathcal{Y}''
-2\mathcal{P}\mathcal{W}^3\mathcal{Y}\ddot{\mathcal{Y}}-\mathcal{P}\mathcal{W}^2\mathcal{Y}^2\ddot{\mathcal{W}}
+\mathcal{P}^2\mathcal{W}\mathcal{Y}^2\mathcal{P}''\\\nonumber
&+\mathcal{P}^3\mathcal{W}\mathcal{Y}'^2-2\mathcal{P}^3\mathcal{Y}\mathcal{W}'\mathcal{Y}'
+2\mathcal{P}^2\mathcal{W}\mathcal{Y}\mathcal{P}'\mathcal{Y}'-\mathcal{P}\mathcal{W}^3\dot{\mathcal{W}}^2
-2\mathcal{P}\mathcal{W}^2\mathcal{Y}\dot{\mathcal{W}}\dot{\mathcal{Y}}\\\nonumber
&+2\mathcal{W}^3\mathcal{Y}\dot{\mathcal{P}}\dot{\mathcal{Y}}-\mathcal{P}^2\mathcal{Y}^2\mathcal{P}'\mathcal{W}'
+\mathcal{W}^2\mathcal{Y}^2\dot{\mathcal{P}}\dot{\mathcal{W}}-\mathcal{P}^3\mathcal{W}^3\bigg],\\\nonumber
\mathcal{T}&=-\mu+\mathrm{P}_{\mathrm{r}}+2\mathrm{P}_{\bot}-3\zeta\Theta,\\\nonumber
\mathcal{Q}&=-\frac{1}{\mathcal{P}^3\mathcal{W}^3\mathcal{Y}}\bigg[\mu\big\{\mathcal{P}\mathcal{W}^2\mathcal{Y}\ddot{\mathcal{W}}
-\mathcal{P}^2\mathcal{W}\mathcal{C}\mathcal{P}''+\mathcal{P}^2\mathcal{YP}'\mathcal{W}'-2\mathcal{P}^2\mathcal{WP}'\mathcal{Y}'\\\nonumber
&+2\mathcal{PW}^3\ddot{\mathcal{Y}}-\mathcal{W}^2\mathcal{Y}\dot{\mathcal{P}}\dot{\mathcal{W}}
-2\mathcal{W}^3\dot{\mathcal{P}}\dot{\mathcal{Y}}\big\}+\big(\mathrm{P}_{\mathrm{r}}-\zeta\Theta\big)
\big\{\mathcal{W}^2\mathcal{Y}\dot{\mathcal{P}}\dot{\mathcal{W}}\\\nonumber
&-\mathcal{PW}^2\mathcal{Y}\ddot{\mathcal{W}}+\mathcal{P}^2\mathcal{WYP}''-2\mathcal{PW}^2\dot{\mathcal{W}}\dot{\mathcal{Y}}
-2\mathcal{P}^3\mathcal{W}'\mathcal{Y}'-\mathcal{P}^2\mathcal{YP}'\mathcal{W}'\\\nonumber
&+2\mathcal{P}^3\mathcal{WY}''\big\}+2\big(\mathrm{P}_{\bot}-\zeta\Theta\big)
\bigg\{\mathcal{P}^3\mathcal{WY}''-\mathcal{P}^3\mathcal{W}'\mathcal{Y}'+\mathcal{P}^2\mathcal{WP}'\mathcal{Y}'\\\nonumber
&-\mathcal{PW}^2\dot{\mathcal{W}}\dot{\mathcal{Y}}-\mathcal{PW}^3\ddot{\mathcal{Y}}+\mathcal{W}^3\dot{\mathcal{P}}\dot{\mathcal{Y}}
-\mathcal{P}^3\mathcal{Y}^3+\frac{\mathcal{P}^3\mathcal{W}\mathcal{Y}'^2}{\mathcal{Y}}
-\frac{\mathcal{P}\mathcal{W}^3\dot{\mathcal{Y}^2}}{\mathcal{Y}}\bigg\}\\\nonumber
&+4\varsigma\mathcal{PW}\big\{\mathcal{PW}\dot{\mathcal{Y}}'
-\mathcal{WP}'\dot{\mathcal{Y}}-\mathcal{P}\dot{\mathcal{W}}\mathcal{Y}'\big\}\bigg],
\end{align}
where $.=\frac{\partial}{\partial t}$. The modified field equations
analogous to the spherical geometry are
\begin{align}\nonumber
\mathcal{T}_{00}^{(\mathrm{EFF})}&=\frac{1}{1-\frac{\Psi{s^2}}{2\mathcal{Y}^4}}\left(8\pi\mu+\frac{s^2}{\mathcal{Y}^4}
+\frac{\mu^{(D)}}{\mathcal{P}^{2}}+\frac{\mathcal{E}_{00}^{(D)}}{\mathcal{P}^{2}}\right)=\frac{\dot{\mathcal{Y}}}{\mathcal{P}^2\mathcal{Y}}
\left(\frac{2\dot{\mathcal{W}}}{\mathcal{W}}+\frac{\dot{\mathcal{Y}}}{\mathcal{Y}}\right)\\\label{g8}
&-\frac{1}{\mathcal{W}^2}\left[\frac{2\mathcal{Y}''}{\mathcal{Y}}+\left(\frac{\mathcal{Y}'}{\mathcal{Y}}\right)^2
-\frac{2\mathcal{W}'\mathcal{Y}'}{\mathcal{WY}}-\left(\frac{\mathcal{W}}{\mathcal{Y}}\right)^2\right],\\\nonumber
\mathcal{T}_{11}^{(\mathrm{EFF})}&=\frac{1}{1-\frac{\Psi{s^2}}{2\mathcal{Y}^4}}\left(8\pi\mathrm{P}_{\mathrm{r}}
-\frac{s^2}{\mathcal{Y}^4}-\zeta\Theta+\frac{\mathrm{P}_{\mathrm{r}}^{(D)}}{\mathcal{W}^{2}}
+\frac{\mathcal{E}_{11}^{(D)}}{\mathcal{W}^{2}}\right)=-\frac{1}{\mathcal{Y}^2}\\\label{g8a}
&+\frac{\mathcal{Y}'}{\mathcal{W}^2\mathcal{Y}}\left(\frac{2\mathcal{P}'}{\mathcal{P}}+\frac{\mathcal{Y}'}{\mathcal{Y}}\right)
-\frac{1}{\mathcal{P}^2}\left[2\frac{\ddot{\mathcal{Y}}}{\mathcal{Y}}-\left(\frac{2\dot{\mathcal{P}}}{\mathcal{P}}
-\frac{\dot{\mathcal{Y}}}{\mathcal{Y}}\right)\frac{\dot{\mathcal{Y}}}{\mathcal{Y}}\right],\\\nonumber
\mathcal{T}_{22}^{(\mathrm{EFF})}&=\frac{\mathcal{T}_{33}^{(\mathrm{EFF})}}{sin^2\theta}=\frac{1}{1-\frac{\Psi{s^2}}{2\mathcal{Y}^4}}
\left(8\pi\mathrm{P}_{\bot}+\frac{s^2}{\mathcal{Y}^4}-\zeta\Theta
+\frac{\mathrm{P}_{\bot}^{(D)}}{\mathcal{Y}^2}\right)=-\frac{1}{\mathcal{P}^2}\\\nonumber
&\times\left[\frac{\ddot{\mathcal{W}}}{\mathcal{W}}+\frac{\ddot{\mathcal{Y}}}{\mathcal{Y}}-\frac{\dot{\mathcal{P}}}{\mathcal{P}}
\left(\frac{\dot{\mathcal{W}}}{\mathcal{W}}+\frac{\dot{\mathcal{Y}}}{\mathcal{Y}}\right)
+\frac{\dot{\mathcal{W}}\dot{\mathcal{Y}}}{\mathcal{WY}}\right]+\frac{1}{\mathcal{W}^2}\\\label{g8c}
&\times\left[\frac{\mathcal{P}''}{\mathcal{P}}+\frac{\mathcal{Y}''}{\mathcal{Y}}-\frac{\mathcal{P}'\mathcal{W}'}{\mathcal{PW}}
+\left(\frac{\mathcal{P}'}{\mathcal{P}}-\frac{\mathcal{W}'}{\mathcal{W}}\right)\frac{\mathcal{Y}'}{\mathcal{Y}}\right],\\\label{g8d}
\mathcal{T}_{01}^{(\mathrm{EFF})}&=\frac{1}{1-\frac{\Psi{s^2}}{2\mathcal{Y}^4}}\left(8\pi\varsigma-\frac{\varsigma^{(D)}}{\mathcal{PW}}
-\frac{\mathcal{E}_{01}^{(D)}}{\mathcal{PW}}\right)=\frac{2}{\mathcal{PW}}\bigg(\frac{\dot{\mathcal{Y}}'}{\mathcal{Y}}
-\frac{\dot{\mathcal{W}}\mathcal{Y}'}{\mathcal{PW}}-\frac{\mathcal{P}'\dot{\mathcal{Y}}}{\mathcal{P}\mathcal{Y}}\bigg),
\end{align}
which couple the fluid distribution with gravity. The entities
$\mu^{(D)}$, $\mathrm{P}_{\mathrm{r}}^{(D)}$,
$\mathrm{P}_{\bot}^{(D)}$, $\varsigma^{(D)}$,
$\mathcal{E}_{00}^{(D)}$, $\mathcal{E}_{11}^{(D)}$ and
$\mathcal{E}_{01}^{(D)}$ in the above equations are due to the
extended gravity. We provide their values in Appendix \textbf{A}.
The other quantities such as the effective energy density, radial
and tangential pressures, and effective heat flux are demonstrated
by
$\bigg(8\pi\mu+\frac{s^2}{\mathcal{Y}^4}+\frac{\mu^{(D)}}{\mathcal{P}^{2}}
+\frac{\mathcal{E}_{00}^{(D)}}{\mathcal{P}^{2}}\bigg)$,
$\bigg(8\pi\mathrm{P}_{\mathrm{r}}-\frac{s^2}{\mathcal{Y}^4}-\zeta\Theta
+\frac{\mathrm{P}_{\mathrm{r}}^{(D)}}{\mathcal{W}^{2}}+\frac{\mathcal{E}_{11}^{(D)}}{\mathcal{W}^{2}}\bigg)$,
$\bigg(8\pi\mathrm{P}_{\bot}+\frac{s^2}{\mathcal{Y}^4}-\zeta\Theta
+\frac{\mathrm{P}_{\bot}^{(D)}}{\mathcal{Y}^2}\bigg)$ and
$\bigg(8\pi\varsigma-\frac{\varsigma^{(D)}}{\mathcal{PW}}
-\frac{\mathcal{E}_{01}^{(D)}}{\mathcal{PW}}\bigg)$.

Misner and Sharp \cite{41bb} provides the formula to calculate mass
of the inner distribution as
\begin{equation}\label{g13}
\tilde{m}(t,r)=\frac{\mathcal{Y}}{2}\left(1-g^{\varphi\vartheta}\partial_{\varphi}\mathcal{Y}\partial_{\vartheta}\mathcal{Y}\right),
\end{equation}
which yields in the presence of charge after some manipulation as
\begin{equation}\label{g14}
\tilde{m}(t,r)=\frac{\mathcal{Y}}{2}\bigg[1-\bigg(\frac{\mathcal{Y}'}{\mathcal{W}}\bigg)^2+\bigg(\frac{\dot{\mathcal{Y}}}{\mathcal{P}}\bigg)^2
+\frac{s^2}{\mathcal{Y}^2}\bigg].
\end{equation}
The inner and outer geometries can be split through a hypersurface
$\Sigma$ at which the solutions to the field equations explain the
spacetime properly. For this, the exterior spacetime is defined as
\begin{equation}\label{g14a}
ds^{2}=-\left(1-\frac{2\mathcal{M}(v)}{R}+\frac{\mathcal{S}^2(v)}{R^2}\right)dv^{2}-2dvdR+R^{2}\left(d\theta^{2}+\sin^{2}\theta
d\phi^{2}\right),
\end{equation}
where $v,~\mathcal{M},~\mathcal{S}$ and $R$ are the retarded time,
mass, charge and radius of the outer region. The Darmois junction
conditions provide the following constraints as
\begin{align}\label{g14b}
&\mathcal{M}{^\Sigma_=}\tilde{m} \quad \Leftrightarrow \quad s
{^\Sigma_=} \mathcal{S}, \\\label{g14c}
&8\pi\mathrm{P}_{\mathrm{r}}-\frac{s^2}{\mathcal{Y}^4}-\zeta\Theta
+\frac{\mathrm{P}_{\mathrm{r}}^{(D)}}{\mathcal{W}^{2}}+\frac{\mathcal{E}_{11}^{(D)}}{\mathcal{W}^{2}}
{^\Sigma_=} 8\pi\varsigma-\frac{\varsigma^{(D)}}{\mathcal{PW}}
-\frac{\mathcal{E}_{01}^{(D)}}{\mathcal{PW}}.
\end{align}
It is interesting to note that the mass and charge of inner and
outer regions of spherical geometry are the same at the
hypersurface, opposing that of cylindrical structure. Equation
\eqref{g14c} demonstrates that the smooth matching results in the
equivalence of effective heat flux and effective radial pressure at
$\Sigma$. The disappearance of radial pressure at the boundary can
be observed only when we exclude the effects of heat flux as well as
modified corrections, i.e.,
$8\pi\varsigma-\frac{\varsigma^{(D)}}{\mathcal{PW}}-\frac{\mathcal{E}_{01}^{(D)}}{\mathcal{PW}}=0$.

\section{Dynamics of the Spherical Star}

Various dynamical entities have been formulated by Misner and Sharp
to study the evolution of spherical structure. The proper temporal
and radial derivatives were employed to determine the velocity and
acceleration of the considered collapsing source. The patterns of
evolution corresponding to spherical/cylindrical spacetimes have
been studied by employing these definitions \cite{41bb}. The
presence of charge in modified scenario leads to the following
equations
\begin{align}\label{g21}
\mathcal{T}_{\varphi;\vartheta}^{(\mathrm{EFF})\vartheta}\mathcal{U}^{\varphi}&=\big\{8\pi\big(\mathcal{T}_{\varphi}^{\vartheta}
+\mathcal{E}_{\varphi}^{\vartheta}\big)+\mathcal{T}_{\varphi}^{(D)\vartheta}\big\}_{;\vartheta}\mathcal{U}^{\varphi}=0,\\\label{g22}
\mathcal{T}_{\varphi;\vartheta}^{(\mathrm{EFF})\vartheta}\mathcal{K}^{\varphi}&=\big\{8\pi\big(\mathcal{T}_{\varphi}^{\vartheta}
+\mathcal{E}_{\varphi}^{\vartheta}\big)+\mathcal{T}_{\varphi}^{(D)\vartheta}\big\}_{;\vartheta}\mathcal{K}^{\varphi}=0,
\end{align}
which yield, respectively, as
\begin{align}\nonumber
&\frac{1}{\mathcal{A}^2}\bigg(8\pi\mu+\frac{s^2}{\mathcal{Y}^4}+\frac{\mu^{(D)}}{\mathcal{P}^{2}}
+\frac{\mathcal{E}_{00}^{(D)}}{\mathcal{P}^{2}}\bigg)^.+\frac{\dot{\mathcal{W}}}{\mathcal{P}^2\mathcal{W}}
\bigg(8\pi\mu+\frac{\mu^{(D)}}{\mathcal{P}^{2}}+\frac{\mathcal{E}_{00}^{(D)}}{\mathcal{P}^{2}}+8\pi\mathrm{P}_{\mathrm{r}}\\\nonumber
&-\zeta\Theta+\frac{\mathrm{P}_{\mathrm{r}}^{(D)}}{\mathcal{W}^{2}}+\frac{\mathcal{E}_{11}^{(D)}}{\mathcal{W}^{2}}\bigg)
+\frac{1}{\mathcal{PW}}\bigg(8\pi\varsigma-\frac{\varsigma^{(D)}}{\mathcal{PW}}-\frac{\mathcal{E}_{01}^{(D)}}{\mathcal{PW}}\bigg)'
+\frac{2\dot{\mathcal{Y}}}{\mathcal{P}^2\mathcal{Y}}\\\nonumber
&\times\bigg(8\pi\mu+\frac{2s^2}{\mathcal{Y}^4}+\frac{\mu^{(D)}}{\mathcal{P}^{2}}+\frac{\mathcal{E}_{00}^{(D)}}{\mathcal{P}^{2}}
+8\pi\mathrm{P}_{\bot}-\zeta\Theta+\frac{\mathrm{P}_{\phi}^{(D)}}{\mathcal{Y}^{2}}\bigg)+\frac{1}{\mathcal{PW}}\\\label{g23}
&\times\bigg(8\pi\varsigma-\frac{\varsigma^{(D)}}{\mathcal{PW}}-\frac{\mathcal{E}_{01}^{(D)}}{\mathcal{PW}}\bigg)
\bigg(\frac{2\mathcal{P}'}{\mathcal{P}}+\frac{\mathcal{Y}'}{\mathcal{Y}}\bigg)=0,\\\nonumber
&\frac{\mathcal{W}}{\mathcal{P}}\bigg(8\pi\varsigma-\frac{\varsigma^{(D)}}{\mathcal{PW}}-\frac{\mathcal{E}_{01}^{(D)}}{\mathcal{PW}}\bigg)^.
+\bigg(8\pi\mathrm{P}_{\mathrm{r}}-\frac{s^2}{\mathcal{Y}^4}-\zeta\Theta+\frac{\mathrm{P}_{\mathrm{r}}^{(D)}}{\mathcal{W}^{2}}
+\frac{\mathcal{E}_{11}^{(D)}}{\mathcal{W}^{2}}\bigg)'\\\nonumber
&+\frac{\mathcal{P}'}{\mathcal{P}}\bigg(8\pi\mu+\frac{\mu^{(D)}}{\mathcal{P}^{2}}+\frac{\mathcal{E}_{00}^{(D)}}{\mathcal{P}^{2}}
+8\pi\mathrm{P}_{\mathrm{r}}-\zeta\Theta+\frac{\mathrm{P}_{\mathrm{r}}^{(D)}}{\mathcal{W}^{2}}
+\frac{\mathcal{E}_{11}^{(D)}}{\mathcal{W}^{2}}\bigg)+\frac{2\mathcal{Y}'}{\mathcal{Y}}\\\nonumber
&\times\bigg(8\pi\mathrm{P}_{\mathrm{r}}-\frac{2s^2}{\mathcal{Y}^4}+\frac{\mathrm{P}_{\mathrm{r}}^{(D)}}{\mathcal{W}^{2}}
+\frac{\mathcal{E}_{11}^{(D)}}{\mathcal{W}^{2}}-8\pi\mathrm{P}_{\bot}-\frac{\mathrm{P}_{\bot}^{(D)}}{\mathcal{Y}^{2}}\bigg)
+\frac{2\mathcal{W}}{\mathcal{P}}\bigg(\frac{\dot{\mathcal{W}}}{\mathcal{W}}+\frac{\dot{\mathcal{Y}}}{\mathcal{Y}}\bigg)\\\label{g24}
&\times\bigg(8\pi\varsigma-\frac{\varsigma^{(D)}}{\mathcal{PW}}-\frac{\mathcal{E}_{01}^{(D)}}{\mathcal{PW}}\bigg)=0.
\end{align}
The above identities play a significant role in examining different
variations taking place in the evolutionary pattern of
self-gravitating structure. The dynamics of collapsing setup can be
well understood by introducing some influential terms such as the
proper radial as well as temporal derivatives defined as
\cite{29a,41bb}
\begin{equation}\label{g25}
\mathfrak{D}_{\mathrm{r}}=\frac{1}{\mathcal{Y}'}
\frac{\partial}{\partial \mathrm{r}}, \quad
\mathfrak{D}_{\mathrm{t}}=\frac{1}{\mathcal{P}}
\frac{\partial}{\partial \mathrm{t}}.
\end{equation}

A compact star remains in the hydrostatic equilibrium until its
gravitational attraction and pressure acting in the outward
direction counterbalances each other. At some point, that pressure
is no more able to produce counter force to gravity, and the star
then collapses. The gradual reduction in radius consequently results
in negative velocity of the interior configuration, i.e.,
\begin{equation}\label{g26}
\mathbb{U}=\mathfrak{D}_{\mathrm{t}}(\mathcal{Y})=\frac{\dot{\mathcal{Y}}}{\mathcal{P}}<0.
\end{equation}
Using this with the mass function \eqref{g14}, we have
\begin{equation}\label{g27}
\frac{\mathcal{Y}'}{\mathcal{W}}=\left(1+\mathbb{U}^{2}-\frac{2\tilde{m}}{\mathcal{Y}}+\frac{s^2}{\mathcal{Y}^2}\right)^{\frac{1}{2}}=\omega.
\end{equation}
The mass of spherical geometry becomes after applying the definition
of $\mathfrak{D}_{\mathrm{t}}$ \eqref{g25} as
\begin{align}\nonumber
\mathfrak{D}_{\mathrm{t}}(\tilde{m})&=-\frac{\mathcal{C}^2}{2\big(1-\frac{\Psi{s^2}}{2\mathcal{Y}^4}\big)}
\left\{\left(8\pi\mathrm{P}_{\mathrm{r}}-\frac{s^2}{\mathcal{Y}^4}-\zeta\Theta
+\frac{\mathrm{P}_{\mathrm{r}}^{(D)}}{\mathcal{W}^{2}}+\frac{\mathcal{E}_{11}^{(D)}}{\mathcal{W}^{2}}\right)\mathbb{U}\right.\\\label{g28}
&\left.+\left(8\pi\varsigma-\frac{\varsigma^{(D)}}{\mathcal{PW}}-\frac{\mathcal{E}_{01}^{(D)}}{\mathcal{PW}}\right)\omega\right\}
-\frac{s^2\dot{\mathcal{Y}}}{2\mathcal{PY}^2},
\end{align}
which demonstrates the fluctuations in total energy of the sphere
with respect to time. Equation \eqref{g28} also discusses the impact
of effective physical terms (radial pressure and heat flux), the
expansion scalar and the electromagnetic field on the collapsing
phenomenon. One can observe that the total energy increases as the
factor
$\left(8\pi\mathrm{P}_{\mathrm{r}}-\frac{s^2}{\mathcal{Y}^4}-\zeta\Theta
+\frac{\mathrm{P}_{\mathrm{r}}^{(D)}}{\mathcal{W}^{2}}+\frac{\mathcal{E}_{11}^{(D)}}{\mathcal{W}^{2}}\right)\mathbb{U}$
(appearing on right side of the above equation) becomes positive due
to negative velocity, and the inequality
$8\pi\mathrm{P}_{\mathrm{r}}-\zeta\Theta+\frac{\mathrm{P}_{\mathrm{r}}^{(D)}}{\mathcal{W}^{2}}>
\frac{s^2}{\mathcal{Y}^4}-\frac{\mathcal{E}_{11}^{(D)}}{\mathcal{W}^{2}}$
exists. This means that the effect of charge is smaller than that of
effective radial pressure acting outward. The term
$\left(8\pi\varsigma-\frac{\varsigma^{(D)}}{\mathcal{AB}}-\frac{\mathcal{E}_{01}^{(D)}}{\mathcal{AB}}\right)\omega$
indicates that the total energy reduces. The last term yields
positive sign and represents Coulomb force that also decreases the
energy of sphere.

The energy between the adjoining spherical surfaces varies. To
discuss this, we implement the definition of
$\mathfrak{D}_{\mathrm{r}}$ on Eq.\eqref{g14} and use it with
Eqs.\eqref{g8} and \eqref{g8d} as
\begin{align}\nonumber
\mathfrak{D}_{\mathrm{r}}(\tilde{m})&=\frac{\mathcal{Y}^2}{2\big(1-\frac{\Psi{s^2}}{2\mathcal{Y}^4}\big)}\left\{\left(8\pi\mu
+\frac{s^2}{\mathcal{Y}^4}+\frac{\mu^{(D)}}{\mathcal{P}^{2}}+\frac{\mathcal{E}_{00}^{(D)}}{\mathcal{P}^{2}}\right)\right.\\\label{g29}
&+\left.\left(8\pi\varsigma-\frac{\varsigma^{(D)}}{\mathcal{PW}}-\frac{\mathcal{E}_{01}^{(D)}}{\mathcal{PW}}\right)
\frac{\mathbb{U}}{\omega}\right\}+\frac{s}{\mathcal{Y}}\mathfrak{D}_{\mathrm{r}}(s)-\frac{s^2}{2\mathcal{Y}^2}.
\end{align}
The quantity
$\left(8\pi\mu+\frac{s^2}{\mathcal{Y}^4}+\frac{\mu^{(D)}}{\mathcal{P}^{2}}+\frac{\mathcal{E}_{00}^{(D)}}{\mathcal{P}^{2}}\right)$
explains how the effective energy density and electric field
intensity affect the collapse rate of the spherical matter source.
The increment in the total energy is mainly controlled by the energy
density. The second term
$\left(8\pi\varsigma-\frac{\varsigma^{(D)}}{\mathcal{PW}}-\frac{\mathcal{E}_{01}^{(D)}}{\mathcal{PW}}\right)\frac{\mathbb{U}}{\omega}$
along with Eq.\eqref{g26} guarantees the dissipation of heat from
the geometry. Also, the impact of charge and Coulomb force can be
observed by $\frac{s}{\mathcal{Y}}\mathfrak{D}_{\mathrm{r}}(s)$ and
$\frac{s^2}{2\mathcal{Y}^2}$. The acceleration of the collapsing
source can be calculated by joining Eqs.\eqref{g25} and \eqref{g26}
as
\begin{align}\nonumber
\mathfrak{D}_{\mathrm{t}}(\mathbb{U})&=-\frac{\tilde{m}}{\mathcal{Y}^2}+\frac{s^2}{2\mathcal{Y}^3}+\frac{\omega\mathcal{P}'}{\mathcal{PW}}
-\frac{\mathcal{C}}{2\big(1-\frac{\Psi{s^2}}{2\mathcal{Y}^4}\big)}\\\label{g30}
&\times\bigg(8\pi\mathrm{P}_{\mathrm{r}}-\frac{s^2}{\mathcal{Y}^4}-\zeta\Theta+\frac{\mathrm{P}_{\mathrm{r}}^{(D)}}{\mathcal{W}^{2}}
+\frac{\mathcal{E}_{11}^{(D)}}{\mathcal{W}^{2}}\bigg).
\end{align}
Equation \eqref{g24} yields the value of
$\frac{\mathcal{P}'}{\mathcal{P}}$ as
\begin{align}\nonumber
\frac{\mathcal{P}'}{\mathcal{P}}&=\frac{-1}{\left(8\pi\mu+\frac{\mu^{(D)}}{\mathcal{P}^{2}}+\frac{\mathcal{E}_{00}^{(D)}}{\mathcal{P}^{2}}
+8\pi\mathrm{P}_{\mathrm{r}}-\zeta\Theta
+\frac{\mathrm{P}_{\mathrm{r}}^{(D)}}{\mathcal{W}^{2}}+\frac{\mathcal{E}_{11}^{(D)}}{\mathcal{W}^{2}}\right)}
\bigg\{\frac{2\mathcal{W}}{\mathcal{P}}\bigg(\frac{\dot{\mathcal{W}}}{\mathcal{W}}+\frac{\dot{\mathcal{Y}}}{\mathcal{Y}}\bigg)\\\nonumber
&\times\bigg(8\pi\varsigma-\frac{\varsigma^{(D)}}{\mathcal{PW}}-\frac{\mathcal{E}_{01}^{(D)}}{\mathcal{PW}}\bigg)
+\bigg(8\pi\mathrm{P}_{\mathrm{r}}-\frac{s^2}{\mathcal{Y}^4}-\zeta\Theta
+\frac{\mathrm{P}_{\mathrm{r}}^{(D)}}{\mathcal{W}^{2}}+\frac{\mathcal{E}_{11}^{(D)}}{\mathcal{W}^{2}}\bigg)'\\\nonumber
&+\frac{2\mathcal{Y}'}{\mathcal{Y}}\bigg(8\pi\mathrm{P}_{\mathrm{r}}-\frac{2s^2}{\mathcal{Y}^4}
+\frac{\mathrm{P}_{\mathrm{r}}^{(D)}}{\mathcal{W}^{2}}+\frac{\mathcal{E}_{11}^{(D)}}{\mathcal{W}^{2}}
-8\pi\mathrm{P}_{\bot}-\frac{\mathrm{P}_{\bot}^{(D)}}{\mathcal{Y}^{2}}\bigg)+\frac{\mathcal{W}}{\mathcal{P}}\\\label{g31}
&\times\bigg(8\pi\varsigma-\frac{\varsigma^{(D)}}{\mathcal{PW}}
-\frac{\mathcal{E}_{01}^{(D)}}{\mathcal{PW}}\bigg)^.\bigg\},
\end{align}
whose insertion in Eq.\eqref{g30} leads to
\begin{align}\nonumber
&\mathfrak{D}_{\mathrm{t}}(\mathbb{U})\bigg(8\pi\mu+\frac{\mu^{(D)}}{\mathcal{P}^{2}}+\frac{\mathcal{E}_{00}^{(D)}}{\mathcal{P}^{2}}
+8\pi\mathrm{P}_{\mathrm{r}}-\zeta\Theta+\frac{\mathrm{P}_{\mathrm{r}}^{(D)}}{\mathcal{W}^{2}}
+\frac{\mathcal{E}_{11}^{(D)}}{\mathcal{W}^{2}}\bigg)=-\bigg\{\frac{\tilde{m}}{\mathcal{Y}^{2}}\\\nonumber
&-\frac{s^2}{2\mathcal{Y}^3}+\frac{\mathcal{Y}}{2\big(1-\frac{\Psi{s^2}}{2\mathcal{Y}^4}\big)}\bigg(8\pi\mathrm{P}_{\mathrm{r}}
-\frac{s^2}{\mathcal{Y}^4}-\zeta\Theta+\frac{\mathrm{P}_{\mathrm{r}}^{(D)}}{\mathcal{W}^{2}}
+\frac{\mathcal{E}_{11}^{(D)}}{\mathcal{W}^{2}}\bigg)\bigg\}\\\nonumber
&\times\bigg(8\pi\mu+\frac{\mu^{(D)}}{\mathcal{P}^{2}}+\frac{\mathcal{E}_{00}^{(D)}}{\mathcal{P}^{2}}
+8\pi\mathrm{P}_{\mathrm{r}}-\zeta\Theta+\frac{\mathrm{P}_{\mathrm{r}}^{(D)}}{\mathcal{W}^{2}}
+\frac{\mathcal{E}_{11}^{(D)}}{\mathcal{W}^{2}}\bigg)-\frac{\omega^{2}}{\mathcal{Y}}\\\nonumber
&\times\bigg(8\pi\mathrm{P}_{\mathrm{r}}-\frac{2s^2}{\mathcal{Y}^4}+\frac{\mathrm{P}_{\mathrm{r}}^{(D)}}{\mathcal{W}^{2}}
+\frac{\mathcal{E}_{11}^{(D)}}{\mathcal{W}^{2}}-8\pi\mathrm{P}_{\bot}-\frac{\mathrm{P}_{\bot}^{(D)}}{\mathcal{Y}^{2}}\bigg)
-\omega\bigg\{\frac{1}{\mathcal{W}}\\\nonumber
&\times\bigg(8\pi\mathrm{P}_{\mathrm{r}}-\frac{s^2}{\mathcal{Y}^4}-\zeta\Theta
+\frac{\mathrm{P}_{\mathrm{r}}^{(D)}}{\mathcal{W}^{2}}+\frac{\mathcal{E}_{11}^{(D)}}{\mathcal{W}^{2}}\bigg)'
+\mathfrak{D}_{\mathrm{t}}\bigg(8\pi\varsigma-\frac{\varsigma^{(D)}}{\mathcal{PW}}-\frac{\mathcal{E}_{01}^{(D)}}{\mathcal{PW}}\bigg)\\\label{g32}
&+\frac{2}{\mathcal{P}}\bigg(8\pi\varsigma-\frac{\varsigma^{(D)}}{\mathcal{PW}}-\frac{\mathcal{E}_{01}^{(D)}}{\mathcal{PW}}\bigg)
\bigg(\frac{\dot{\mathcal{W}}}{\mathcal{W}}+\frac{\dot{\mathcal{Y}}}{\mathcal{Y}}\bigg)\bigg\}.
\end{align}
The left side of Eq.\eqref{g32} represents the Newtonian force as
the product of acceleration ($\mathfrak{D}_{\mathrm{t}}\mathbb{U}$)
and the inertial mass density
$\bigg(8\pi\mu+\frac{\mu^{(D)}}{\mathcal{P}^{2}}+\frac{\mathcal{E}_{00}^{(D)}}{\mathcal{P}^{2}}
+8\pi\mathrm{P}_{\mathrm{r}}-\zeta\Theta+\frac{\mathrm{P}_{\mathrm{r}}^{(D)}}{\mathcal{W}^{2}}
+\frac{\mathcal{E}_{11}^{(D)}}{\mathcal{W}^{2}}\bigg)$ arises. The
later term also appears on right side at the same time, now offering
the gravitational mass density. The indistinguishability between
both of these masses confirms the satisfaction of the equivalence
principle. The expression with $\omega^2$ in small bracket
determines the effect of state variables along with correction terms
on the collapse rate. The role of gradient of modified radial
pressure in this scenario can also be observed through first term in
the last curly bracket, which results in the reduction of collapse
rate. Moreover, the last two factors involving effective heat flux
and its time derivative well describe the hydrodynamical features of
the charged sphere.

\section{Transport Equations}

The structural changes of any self-gravitating body can
substantially be discussed with the help of some dynamical
equations. The transport equation is one of them which plays highly
significant role in discussing the sphere coupled with
$\mathbb{EMT}$ \eqref{g5}. Such an equation helps to evaluate
certain physical quantities (like mass, momentum and heat) during
the collapse. This equation also supports the diffusion process
inside the spherical object and has the following form in modified
framework as
\begin{equation}\label{g33}
\varrho\mathrm{h}^{\varphi\vartheta}\mathcal{U}^{\gamma}\bar{\varsigma}_{\vartheta;\gamma}+\bar{\varsigma}^{\varphi}=
-\eta\mathrm{h}^{\varphi\vartheta}\left(\tau_{,\vartheta}+\tau\mathrm{a}_{\vartheta}\right)-\frac{1}{2}\eta\tau^{2}\left(\frac{\varrho
\mathcal{U}^{\vartheta}}{\eta\tau^{2}}\right)_{;\vartheta}\bar{\varsigma}^{\varphi},
\end{equation}
where
$\bar{\varsigma}=\left(8\pi\varsigma-\frac{\varsigma^{(D)}}{\mathcal{PW}}-\frac{\mathcal{E}_{01}^{(D)}}{\mathcal{PW}}\right)$
and the projection tensor is defined as
$\mathrm{h}^{\varphi\vartheta}=g^{\varphi\vartheta}+\mathcal{U}^{\varphi}
\mathcal{U}^{\vartheta}$. Also, the temperature, thermal
conductivity, relaxation time and four-acceleration are
mathematically represented by $\tau,~\eta,~\varrho$ and
$\mathrm{a}_{\vartheta}$ \big(i.e.,
$\mathrm{a}_1=\frac{\mathcal{A}'}{\mathcal{A}}$\big), respectively.
After simplifying this equation, it gives
\begin{align}\nonumber
\mathcal{W}
\mathfrak{D}_{\mathrm{t}}\left(8\pi\varsigma-\frac{\varsigma^{(D)}}{\mathcal{PW}}-\frac{\mathcal{E}_{01}^{(D)}}{\mathcal{PW}}\right)&=
-\frac{\eta\tau^{2}\mathcal{W}}{2\mathcal{P}\varrho}\left(\frac{\varrho}{\eta\tau^{2}}\right)^.
\left(8\pi\varsigma-\frac{\varsigma^{(D)}}{\mathcal{PW}}-\frac{\mathcal{E}_{01}^{(D)}}{\mathcal{PW}}\right)\\\nonumber
&-\frac{\eta\tau'}{\varrho}-\frac{\eta\tau}{\varrho}\left(\frac{\mathcal{P}'}{\mathcal{P}}\right)-\frac{\mathcal{W}}{2\mathcal{P}}
\left(\frac{3\dot{\mathcal{W}}}{\mathcal{W}}+\frac{2\dot{\mathcal{Y}}}{\mathcal{Y}}+\frac{2\mathcal{P}}{\varrho}\right)\\\label{g34}
&\times\left(8\pi\varsigma-\frac{\varsigma^{(D)}}{\mathcal{PW}}-\frac{\mathcal{E}_{01}^{(D)}}{\mathcal{PW}}\right),
\end{align}
which after using together with Eq.\eqref{g31} yields
\begin{align}\nonumber
\mathcal{W}
\mathfrak{D}_{\mathrm{t}}\bigg(8\pi\varsigma-\frac{\varsigma^{(D)}}{\mathcal{PW}}-\frac{\mathcal{E}_{01}^{(D)}}{\mathcal{PW}}\bigg)&=
-\frac{\eta\tau^{2}\mathcal{W}}{2\mathcal{P}\varrho}\bigg(\frac{\varrho}{\eta\tau^{2}}\bigg)^.\left(8\pi\varsigma
-\frac{\varsigma^{(D)}}{\mathcal{PW}}-\frac{\mathcal{E}_{01}^{(D)}}{\mathcal{PW}}\right)\\\nonumber
&-\frac{\eta\tau'}{\varrho}-\frac{\eta\tau\mathcal{W}}{\varrho\omega}\bigg\{\mathfrak{D}_{\mathrm{t}}(\mathbb{U})
+\frac{\tilde{m}}{\mathcal{Y}^2}+\frac{\mathcal{Y}}{2\big(1-\frac{\Psi{s^2}}{2\mathcal{Y}^4}\big)}\\\nonumber
&\times\bigg(8\pi\mathrm{P}_{\mathrm{r}}-\frac{s^2}{\mathcal{Y}^4}-\zeta\Theta+\frac{\mathrm{P}_{\mathrm{r}}^{(D)}}{\mathcal{W}^{2}}
+\frac{\mathcal{E}_{11}^{(D)}}{\mathcal{W}^{2}}\bigg)\bigg\}-\frac{s^2}{2\mathcal{Y}^3}\\\label{g35}
&-\frac{\mathcal{W}}{2\mathcal{P}}\bigg(8\pi\varsigma-\frac{\varsigma^{(D)}}{\mathcal{PW}}-\frac{\mathcal{E}_{01}^{(D)}}{\mathcal{PW}}\bigg)
\bigg(\frac{3\dot{\mathcal{W}}}{\mathcal{W}}+\frac{2\dot{\mathcal{Y}}}{\mathcal{Y}}+\frac{2\mathcal{P}}{\varrho}\bigg).
\end{align}
This equation shows the variations in heat dissipation with respect
to time and also reveals the influence of thermal conductivity,
relaxation time and temperature on self-gravitating object.

We now eliminate
$\mathfrak{D}_{\mathrm{t}}\left(8\pi\varsigma-\frac{\varsigma^{(D)}}{\mathcal{PW}}-\frac{\mathcal{E}_{01}^{(D)}}{\mathcal{PW}}\right)$
from Eqs.\eqref{g32} and \eqref{g35} to get
\begin{align}\nonumber
&\mathfrak{D}_{\mathrm{t}}(\mathbb{U})\bigg(8\pi\mu+\frac{\mu^{(D)}}{\mathcal{P}^{2}}+\frac{\mathcal{E}_{00}^{(D)}}{\mathcal{P}^{2}}
+8\pi\mathrm{P}_{\mathrm{r}}-\zeta\Theta+\frac{\mathrm{P}_{\mathrm{r}}^{(D)}}{\mathcal{W}^{2}}
+\frac{\mathcal{E}_{11}^{(D)}}{\mathcal{W}^{2}}-\frac{\eta\tau}{\varrho}\bigg)=-\bigg\{\frac{\tilde{m}}{\mathcal{Y}^{2}}\\\nonumber
&-\frac{s^2}{2\mathcal{Y}^3}+\frac{\mathcal{Y}}{2\big(1-\frac{\Psi{s^2}}{2\mathcal{Y}^4}\big)}
\bigg(8\pi\mathrm{P}_{\mathrm{r}}-\frac{s^2}{\mathcal{Y}^4}-\zeta\Theta+\frac{\mathrm{P}_{\mathrm{r}}^{(D)}}{\mathcal{W}^{2}}
+\frac{\mathcal{E}_{11}^{(D)}}{\mathcal{W}^{2}}\bigg)\bigg\}\\\nonumber
&\times\bigg(8\pi\mu+\frac{\mu^{(D)}}{\mathcal{P}^{2}}+\frac{\mathcal{E}_{00}^{(D)}}{\mathcal{P}^{2}}
+8\pi\mathrm{P}_{\mathrm{r}}-\zeta\Theta+\frac{\mathrm{P}_{\mathrm{r}}^{(D)}}{\mathcal{W}^{2}}
+\frac{\mathcal{E}_{11}^{(D)}}{\mathcal{W}^{2}}-\frac{\eta\tau}{\varrho}\bigg)-\frac{2\omega^{2}}{\mathcal{Y}}\\\nonumber
&\times\bigg(8\pi\mathrm{P}_{\mathrm{r}}-\frac{2s^2}{\mathcal{Y}^4}+\frac{\mathrm{P}_{\mathrm{r}}^{(D)}}{\mathcal{W}^{2}}
+\frac{\mathcal{E}_{11}^{(D)}}{\mathcal{W}^{2}}-8\pi\mathrm{P}_{\bot}-\frac{\mathrm{P}_{\bot}^{(D)}}{\mathcal{Y}^{2}}\bigg)
-\omega\bigg[-\frac{\eta\tau'}{\varrho\mathcal{W}}+\frac{1}{\mathcal{W}}\\\nonumber
&\times\bigg(8\pi\mathrm{P}_{\mathrm{r}}-\frac{s^2}{\mathcal{Y}^4}-\zeta\Theta+\frac{\mathrm{P}_{\mathrm{r}}^{(D)}}{\mathcal{W}^{2}}
+\frac{\mathcal{E}_{11}^{(D)}}{\mathcal{W}^{2}}\bigg)'+\frac{1}{2\mathcal{P}}\bigg(8\pi\varsigma-\frac{\varsigma^{(D)}}{\mathcal{PW}}
-\frac{\mathcal{E}_{01}^{(D)}}{\mathcal{PW}}\bigg)\\\label{g36}
&\times\bigg\{\frac{\dot{\mathcal{W}}}{\mathcal{W}}+\frac{2\dot{\mathcal{Y}}}{\mathcal{Y}}-\frac{2\mathcal{P}}{\varrho}
-\frac{\eta\tau^2}{\varrho}\bigg(\frac{\varrho}{\eta\tau^2}\bigg)^.\bigg\}\bigg],
\end{align}
which can be written in concise form as
\begin{align}\nonumber
\mathcal{F}_{\mathrm{newt}}\big(1-\mathfrak{H}\big)&=\omega\bigg[-\frac{\eta\tau'}{\varrho\mathcal{W}}
+\frac{1}{2\mathcal{P}}\bigg\{\frac{\dot{\mathcal{W}}}{\mathcal{W}}+\frac{2\dot{\mathcal{Y}}}{\mathcal{Y}}-\frac{2\mathcal{P}}{\varrho}
-\frac{\eta\tau^2}{\varrho}\bigg(\frac{\varrho}{\eta\tau^2}\bigg)^.\bigg\}\\\label{g37}
&\times\bigg(8\pi\varsigma-\frac{\varsigma^{(D)}}{\mathcal{PW}}-\frac{\mathcal{E}_{01}^{(D)}}{\mathcal{PW}}\bigg)\bigg]
-\mathcal{F}_{\mathrm{grav}}\big(1-\mathfrak{H}\big)+\mathcal{F}_{\mathrm{hyd}},
\end{align}
with
\begin{align}\label{g38}
\mathfrak{H}=&\frac{\eta\tau}{\varrho}\bigg(8\pi\mu+\frac{\mu^{(D)}}{\mathcal{P}^{2}}+\frac{\mathcal{E}_{00}^{(D)}}{\mathcal{P}^{2}}
+8\pi\mathrm{P}_{\mathrm{r}}-\zeta\Theta+\frac{\mathrm{P}_{\mathrm{r}}^{(D)}}{\mathcal{W}^{2}}
+\frac{\mathcal{E}_{11}^{(D)}}{\mathcal{W}^{2}}\bigg)^{-1},\\\label{g38a}
\mathcal{F}_{\mathrm{newt}}=&\mathfrak{D}_{\mathrm{t}}(\mathbb{U})\bigg(8\pi\mu+\frac{\mu^{(D)}}{\mathcal{P}^{2}}
+\frac{\mathcal{E}_{00}^{(D)}}{\mathcal{P}^{2}}+8\pi\mathrm{P}_{\mathrm{r}}-\zeta\Theta+\frac{\mathrm{P}_{\mathrm{r}}^{(D)}}{\mathcal{W}^{2}}
+\frac{\mathcal{E}_{11}^{(D)}}{\mathcal{W}^{2}}\bigg),\\\nonumber
\mathcal{F}_{\mathrm{grav}}=&\bigg(8\pi\mu+\frac{\mu^{(D)}}{\mathcal{P}^{2}}+\frac{\mathcal{E}_{00}^{(D)}}{\mathcal{P}^{2}}
+8\pi\mathrm{P}_{\mathrm{r}}-\zeta\Theta+\frac{\mathrm{P}_{\mathrm{r}}^{(D)}}{\mathcal{W}^{2}}
+\frac{\mathcal{E}_{11}^{(D)}}{\mathcal{W}^{2}}\bigg)\bigg\{\frac{\tilde{m}}{\mathcal{Y}^{2}}\\\label{g39}
&-\frac{s^2}{2\mathcal{Y}^3}+\frac{\mathcal{Y}}{2\big(1-\frac{\Psi{s^2}}{2\mathcal{Y}^4}\big)}
\bigg(8\pi\mathrm{P}_{\mathrm{r}}-\frac{s^2}{\mathcal{Y}^4}-\zeta\Theta+\frac{\mathrm{P}_{\mathrm{r}}^{(D)}}{\mathcal{W}^{2}}
+\frac{\mathcal{E}_{11}^{(D)}}{\mathcal{W}^{2}}\bigg)\bigg\},\\\nonumber
\mathcal{F}_{\mathrm{hyd}}=&-\omega^2\bigg[\mathfrak{D}_{\mathrm{r}}\bigg(8\pi\mathrm{P}_{\mathrm{r}}-\frac{s^2}{\mathcal{Y}^4}-\zeta\Theta
+\frac{\mathrm{P}_{\mathrm{r}}^{(D)}}{\mathcal{W}^{2}}+\frac{\mathcal{E}_{11}^{(D)}}{\mathcal{W}^{2}}\bigg)\\\label{g40}
&+\frac{2}{\mathcal{Y}}\bigg(8\pi\mathrm{P}_{\mathrm{r}}
-\frac{2s^2}{\mathcal{Y}^4}+\frac{\mathrm{P}_{\mathrm{r}}^{(D)}}{\mathcal{W}^{2}}+\frac{\mathcal{E}_{11}^{(D)}}{\mathcal{W}^{2}}
-8\pi\mathrm{P}_{\bot}-\frac{\mathrm{P}_{\bot}^{(D)}}{\mathcal{Y}^{2}}\bigg)\bigg].
\end{align}

It can be observed from Eq.\eqref{g37} that the fluid undergoing
collapsing phenomenon is affected by several forces, including
Newtonian ($\mathcal{F}_{\mathrm{newt}}$), gravitational
($\mathcal{F}_{\mathrm{grav}}$) and hydrodynamical
($\mathcal{F}_{\mathrm{hyd}}$). The gravitational force \eqref{g39}
expresses that how the collapse rate is influenced from mass
function, effective radial pressure and charge. The first entity in
the hydrodynamical force \eqref{g40} represents the rate of change
in radial pressure, which helps to reduce the rate of collapse. The
second round bracket indicates the effective anisotropy of
collapsing source. It should be acknowledged that the positive (or
negative) value of pressure anisotropy enhances (or diminishes) the
collapse rate.

The first and third terms of Eq.\eqref{g37} involving an entity
$(1-\mathfrak{H})$ guarantees fulfillment of the principle of
equivalence. Equation \eqref{g38} displays the inverse relation of
the gravitational mass density with the term $(\mathfrak{H})$. We
can observe that $1-\mathfrak{H}$ strongly affects the gravitational
force of spherical system which leads to different cases. Multiple
possibilities are provided in the following to discuss increment or
decrement in the collapse rate. These cases are
\begin{itemize}
\item The positive contribution of $f(\mathcal{R},\mathcal{T},\mathcal{Q})$ corrections makes $\mathfrak{H}$ smaller than $1$,
thus $1-\mathfrak{H}$ becomes positive. However, the minus sign
appearing with the gravitational force in Eq.\eqref{g37} produces a
repulsive force that ultimately diminishes the collapse rate.

\item The negative effects of modified terms may yield $\mathfrak{H}>1$, (i.e., $1-\mathfrak{H}<0$). This results in increment in the
gravitational force, and thus the collapse rate increases.

\item Furthermore, the consideration $\mathfrak{H}=1$ disappears the gravitational and inertial forces from Eq.\eqref{g37}, thus
we are left with
\begin{align}\nonumber
&\omega^2\bigg[\mathfrak{D}_{\mathrm{r}}\bigg(8\pi\mathrm{P}_{\mathrm{r}}-\frac{s^2}{\mathcal{Y}^4}-\zeta\Theta
+\frac{\mathrm{P}_{\mathrm{r}}^{(D)}}{\mathcal{W}^{2}}+\frac{\mathcal{E}_{11}^{(D)}}{\mathcal{W}^{2}}\bigg)+\frac{2}{\mathcal{Y}}\\\nonumber
&\times\bigg(8\pi\mathrm{P}_{\mathrm{r}}
-\frac{2s^2}{\mathcal{Y}^4}+\frac{\mathrm{P}_{\mathrm{r}}^{(D)}}{\mathcal{W}^{2}}+\frac{\mathcal{E}_{11}^{(D)}}{\mathcal{W}^{2}}
-8\pi\mathrm{P}_{\bot}-\frac{\mathrm{P}_{\bot}^{(D)}}{\mathcal{Y}^{2}}\bigg)\bigg]=\omega\\\nonumber
&\times\bigg[-\frac{\eta\tau'}{\varrho\mathcal{W}}
+\frac{1}{2\mathcal{P}}\bigg\{\frac{\dot{\mathcal{W}}}{\mathcal{W}}+\frac{2\dot{\mathcal{Y}}}{\mathcal{Y}}-\frac{2\mathcal{P}}{\varrho}
-\frac{\eta\tau^2}{\varrho}\bigg(\frac{\varrho}{\eta\tau^2}\bigg)^.\bigg\}\\\label{g41}
&\times\bigg(8\pi\varsigma-\frac{\varsigma^{(D)}}{\mathcal{PW}}-\frac{\mathcal{E}_{01}^{(D)}}{\mathcal{PW}}\bigg)\bigg],
\end{align}
\end{itemize}
which declares the impact of various quantities such as the
temperature, thermal conductivity and the bulk viscosity together
with modified corrections on spherical collapse. The hydrodynamical
force still exists in the above equation due to which the considered
sphere maintains its equilibrium position, and resultantly the
collapse rate is reduced.

\section{Relation between the Weyl Scalar and Effective Physical Quantities}

In this section, we explore certain significant relations between
the Weyl scalar (defined as
$\mathbb{C}^2=\mathfrak{C}_{\varphi\mu\vartheta\nu}\mathfrak{C}^{\varphi\mu\vartheta\nu}$,
where $\mathfrak{C}_{\varphi\mu\vartheta\nu}$ symbolizes the Weyl
tensor) and effective variables such as the energy density and
pressure components. This scalar can be expressed as a linear
combination of the Ricci tensor ($\mathcal{R}_{\varphi\vartheta}$),
the Ricci scalar and the Kretchmann scalar
($\mathbb{R}=\mathcal{R}_{\varphi\mu\vartheta\nu}\mathcal{R}^{\varphi\mu\vartheta\nu}$)
as \cite{35c}
\begin{align}\label{g42}
\mathbb{C}^2=-\frac{1}{3}\big(6\mathcal{R}_{\varphi\vartheta}\mathcal{R}^{\varphi\vartheta}-3\mathbb{R}-\mathcal{R}^2\big).
\end{align}
The scalar $\mathbb{R}$ is manipulated as
\begin{align}\nonumber
\mathbb{R}&=\frac{48}{\mathcal{Y}^6}\bigg(\tilde{m}-\frac{s^2}{2\mathcal{Y}}\bigg)^2-\frac{16}{\mathcal{Y}^3}\bigg(\tilde{m}
-\frac{s^2}{2\mathcal{Y}}\bigg)\bigg(\frac{\mathcal{G}_{00}}{\mathcal{P}^2}-\frac{\mathcal{G}_{11}}{\mathcal{W}^2}
+\frac{\mathcal{G}_{22}}{\mathcal{Y}^2}\bigg)\\\nonumber
&+4\bigg\{\bigg(\frac{\mathcal{G}_{22}}{\mathcal{Y}^2}\bigg)^2-\bigg(\frac{\mathcal{G}_{01}}{\mathcal{PW}}\bigg)^2\bigg\}
+3\bigg\{\bigg(\frac{\mathcal{G}_{00}}{\mathcal{P}^2}\bigg)^2+\bigg(\frac{\mathcal{G}_{11}}{\mathcal{W}^2}\bigg)^2\bigg\}\\\label{g43}
&-\frac{2\mathcal{G}_{00}\mathcal{G}_{11}}{\mathcal{P}^2\mathcal{W}^2}+\frac{4\mathcal{G}_{22}}{\mathcal{Y}^2}
\bigg(\frac{\mathcal{G}_{00}}{\mathcal{P}^2}-\frac{\mathcal{G}_{11}}{\mathcal{W}^2}\bigg).
\end{align}
The sphere \eqref{g6} yields non-null elements of
$\mathcal{R}_{\varphi\vartheta}$ and
$\mathcal{R}_{\varphi\mu\vartheta\nu}$ whose values are provided in
Appendix \textbf{B}. After using them along with Eq.\eqref{g43} and
the Ricci scalar in \eqref{g42}, we have the Weyl scalar as
\begin{align}\nonumber
\frac{\mathbb{C}\mathcal{Y}^3}{\sqrt{48}}&=\tilde{m}-\frac{\mathcal{Y}^3}{6\big(1-\frac{\Psi{s^2}}{2\mathcal{Y}^4}\big)}
\bigg\{8\pi\mu+\frac{3s^2}{\mathcal{Y}^4}+\frac{\mu^{(D)}}{\mathcal{P}^{2}}+\frac{\mathcal{E}_{00}^{(D)}}{\mathcal{P}^{2}}\\\label{g44}
&-8\pi\mathrm{P}_{\mathrm{r}}-\frac{\mathrm{P}_{\mathrm{r}}^{(D)}}{\mathcal{W}^{2}}-\frac{\mathcal{E}_{11}^{(D)}}{\mathcal{W}^{2}}
+8\pi\mathrm{P}_{\bot}+\frac{\mathrm{P}_{\bot}^{(D)}}{\mathcal{Y}^{2}}\bigg\}-\frac{s^2}{2\mathcal{Y}}.
\end{align}
Applying the proper temporal as well as radial derivatives on
Eq.\eqref{g44} and joining them with
$\mathfrak{D}_{\mathrm{t}}(\tilde{m})$ and
$\mathfrak{D}_{\mathrm{r}}(\tilde{m})$, we, respectively have
\begin{align}\nonumber
\mathfrak{D}_{\mathrm{t}}\bigg(\frac{\mathbb{C}\mathcal{Y}^3}{\sqrt{48}}\bigg)&=-\frac{\mathcal{Y}^2}{2\big(1-\frac{\Psi{s^2}}
{2\mathcal{Y}^4}\big)}\bigg\{\bigg(8\pi\mu+\frac{\mu^{(D)}}{\mathcal{P}^{2}}+\frac{\mathcal{E}_{00}^{(D)}}{\mathcal{P}^{2}}
+8\pi\mathrm{P}_{\mathrm{r}}-\zeta\Theta\\\nonumber
&+\frac{\mathrm{P}_{\mathrm{r}}^{(D)}}{\mathcal{W}^{2}}+\frac{\mathcal{E}_{11}^{(D)}}{\mathcal{W}^{2}}\bigg)\mathbb{U}+\bigg(8\pi\varsigma
-\frac{\varsigma^{(D)}}{\mathcal{PW}}-\frac{\mathcal{E}_{01}^{(D)}}{\mathcal{PW}}\bigg)\omega\bigg\}-\frac{\mathcal{Y}^3}{6}\\\nonumber
&\times\mathfrak{D}_{\mathrm{t}}\bigg\{\frac{1}{1-\frac{\Psi{s^2}}{2\mathcal{Y}^4}}\bigg(8\pi\mu+\frac{3s^2}{\mathcal{Y}^4}
+\frac{\mu^{(D)}}{\mathcal{P}^{2}}+\frac{\mathcal{E}_{00}^{(D)}}{\mathcal{P}^{2}}-8\pi\mathrm{P}_{\mathrm{r}}\\\label{g45}
&-\frac{\mathrm{P}_{\mathrm{r}}^{(D)}}{\mathcal{W}^{2}}-\frac{\mathcal{E}_{11}^{(D)}}{\mathcal{W}^{2}}
+8\pi\mathrm{P}_{\bot}+\frac{\mathrm{P}_{\bot}^{(D)}}{\mathcal{Y}^{2}}\bigg)\bigg\},\\\nonumber
\mathfrak{D}_{\mathrm{r}}\bigg(\frac{\mathbb{C}\mathcal{Y}^3}{\sqrt{48}}\bigg)&=\frac{\mathcal{Y}^2}{2\big(1-\frac{\Psi{s^2}}
{2\mathcal{Y}^4}\big)}\bigg\{8\pi\mathrm{P}_{\mathrm{r}}-\frac{2s^2}{\mathcal{Y}^4}+\frac{\mathrm{P}_{\mathrm{r}}^{(D)}}{\mathcal{W}^{2}}
+\frac{\mathcal{E}_{11}^{(D)}}{\mathcal{W}^{2}}-8\pi\mathrm{P}_{\bot}\\\nonumber
&-\frac{\mathrm{P}_{\bot}^{(D)}}{\mathcal{Y}^2}+\bigg(8\pi\varsigma-\frac{\varsigma^{(D)}}{\mathcal{PW}}
-\frac{\mathcal{E}_{01}^{(D)}}{\mathcal{PW}}\bigg)\frac{\mathbb{U}}{\omega}\bigg\}-\frac{\mathcal{Y}^3}{6}\\\nonumber
&\times\mathfrak{D}_{\mathrm{r}}\bigg\{\frac{1}{1-\frac{\Psi{s^2}}{2\mathcal{Y}^4}}\bigg(8\pi\mu+\frac{3s^2}{\mathcal{Y}^4}
+\frac{\mu^{(D)}}{\mathcal{P}^{2}}+\frac{\mathcal{E}_{00}^{(D)}}{\mathcal{P}^{2}}-8\pi\mathrm{P}_{\mathrm{r}}\\\label{g46}
&-\frac{\mathrm{P}_{\mathrm{r}}^{(D)}}{\mathcal{W}^{2}}-\frac{\mathcal{E}_{11}^{(D)}}{\mathcal{W}^{2}}
+8\pi\mathrm{P}_{\bot}+\frac{\mathrm{P}_{\bot}^{(D)}}{\mathcal{Y}^{2}}\bigg)\bigg\}.
\end{align}
Equation \eqref{g46} takes the form in the absence of charge (i.e.,
$s=0$) as
\begin{align}\nonumber
\mathfrak{D}_{\mathrm{r}}\bigg(\frac{\mathbb{C}\mathcal{Y}^3}{\sqrt{48}}\bigg)&=\frac{\mathcal{Y}^2}{2}\bigg\{8\pi\mathrm{P}_{\mathrm{r}}
+\frac{\mathrm{P}_{\mathrm{r}}^{(D)}}{\mathcal{W}^{2}}-8\pi\mathrm{P}_{\bot}-\frac{\mathrm{P}_{\bot}^{(D)}}{\mathcal{Y}^2}
+\bigg(8\pi\varsigma-\frac{\varsigma^{(D)}}{\mathcal{PW}}\bigg)\frac{\mathbb{U}}{\omega}\bigg\}\\\label{g46a}
&-\frac{\mathcal{Y}^3}{6}\mathfrak{D}_{\mathrm{r}}\bigg\{8\pi\mu+\frac{\mu^{(D)}}{\mathcal{P}^{2}}-8\pi\mathrm{P}_{\mathrm{r}}
-\frac{\mathrm{P}_{\mathrm{r}}^{(D)}}{\mathcal{W}^{2}}+8\pi\mathrm{P}_{\bot}+\frac{\mathrm{P}_{\bot}^{(D)}}{\mathcal{Y}^{2}}\bigg\}.
\end{align}

Di Prisco \emph{et al.} \cite{35c} examined the collapse of
spherical body coupled with anisotropic configuration and concluded
that the homogeneous energy density and conformal flatness of that
object are necessary and sufficient conditions for each other. The
validity of this result can be checked in the present scenario of
modified theory by taking a particular form of the corresponding
generic functional. We consider the standard model \eqref{g5d} in
this regard for our convenience. We further assume
$\mathcal{R}=\mathcal{R}_0$ and treat
$f_2(\mathcal{T}=\mathcal{T}_0)$ as well as
$f_3(\mathcal{Q}=\mathcal{Q}_0)$ as constant terms to make sure
validity of the above result. Consequently, Eq.\eqref{g46a} yields
\begin{eqnarray}\label{g47}
\mathfrak{D}_{\mathrm{r}}\bigg(\frac{\mathbb{C}\mathcal{Y}^3}{\sqrt{48}}\bigg)=4\pi\mathcal{Y}^2\bigg\{\mathrm{P}_{\mathrm{r}}
-\mathrm{P}_{\bot}+\frac{\varsigma\mathbb{U}}{\omega}\bigg\}-\frac{4\pi\mathcal{Y}^3}{3}\mathfrak{D}_{\mathrm{r}}\bigg\{\mu
-\mathrm{P}_{\mathrm{r}}+\mathrm{P}_{\bot}-\frac{\mathcal{C}_{0}}{2}\bigg\}.
\end{eqnarray}
where
$\mathcal{C}_{0}=\Phi\sqrt{\mathcal{T}_{0}}+\Psi\mathcal{Q}_{0}$
indicates a constant. From the above equation, we observe the
presence of energy density inhomogeneity in the considered source
due to the appearance of both the principal pressures and
dissipation flux. This also interprets that inhomogeneity of the
sphere during evolution may be caused by the tidal forces
\cite{36a}. The above condition can only be achieved by neglecting
pressure anisotropy and heat flux, Eq.\eqref{g47} is left with
\begin{eqnarray}\label{g48}
\mathfrak{D}_{\mathrm{r}}\bigg(\frac{\mathbb{C}\mathcal{Y}^3}{\sqrt{48}}\bigg)=-\frac{4\pi\mathcal{Y}^3}{3}
\mathfrak{D}_{\mathrm{r}}\bigg\{\mu-\frac{\mathcal{C}_{0}}{2}\bigg\},
\end{eqnarray}
which confirms that homogeneity in the energy density of sphere
(i.e., $\mathfrak{D}_{\mathrm{r}}(\mu)=0$) and the conformal
flatness condition ($\mathfrak{D}_{\mathrm{r}}(\mathbb{C})=0
\Rightarrow \mathbb{C}=0$ through regular axis condition) implies
each other.

\section{Concluding Remarks}

Our universe is made up of multiple constituents comprising of a
large number of astronomical bodies. The formation of such massive
structures takes place due to a captivating phenomenon, known as the
gravitational collapse. After studying the gravitational waves by
means of many observations (i.e., laser interferometric detectors
like GEO, LIGO, TAMA and VIRGO), several astronomers have been
prompted to investigate the collapse rate of celestial objects in
the context of $\mathbb{GR}$ as well as extended theories \cite{44}.
This article examines the dynamics of spherically symmetric
self-gravitating anisotropic configuration in
$f(\mathcal{R},\mathcal{T},\mathcal{R}_{\phi\psi}\mathcal{T}^{\phi\psi})$
gravity. The sphere \eqref{g6} is considered to be influenced by the
electromagnetic field, principal stresses in radial and tangential
directions (such as $\mathrm{P}_{\mathrm{r}}$ and
$\mathrm{P}_{\bot}$), heat flux and the expansion scalar. We have
then studied multiple substantial parameters that produce
evolutionary changes within the matter source. The relations between
mass and charge of both the exterior and interior regions have been
obtained at the hypersurface through Darmois junction criteria. We
have then employed the formulation presented by Misner and Sharp to
extract the non-vanishing elements of the Bianchi identities. The
variations appearing in total energy of the current setup have been
calculated by taking the proper radial/time derivatives of the
corresponding mass function \eqref{g14}.

Several forces (such as the Newtonian, hydrodynamical and
gravitational) have been identified during the formulation of the
transport equation which are then coupled with dynamical identities
to explore the impact of charge, pressure anisotropy and modified
corrections on the collapse rate. The presence of charge has gained
much importance in the dynamics of collapse which acts as a
repulsive force and thus the process of collapse is resisted. We
summarize our results in the following.
\begin{itemize}
\item The outward directed radial pressure and the Coulomb force (having repulsive
nature) decrease total energy of the spherical source, which is
described by the time variation of mass.

\item The variation of mass in the radial direction manifests that the
effective energy density (and the Coulomb force) increase (and
decrease) total energy of the system.

\item The inertial mass density is significantly affected by the $f(\mathcal{R},\mathcal{T},\mathcal{R}_{\phi\psi}\mathcal{T}^{\phi\psi})$
corrections whose positive values assure decrement in the collapse
rate and negative effects do the opposite.

\item It can be observed that the electromagnetic field produces the repulsive effects due to which the collapse rate is reduced.

\item The hydrodynamical force reveals that the negative impact of pressure gradient and the anisotropy slows down the rate of collapse.

\item The gravitational force in this case of charged fluid
decreases as compared to the uncharged distribution \cite{27aad},
which also diminishes the collapse rate.

\item We have also analyzed from Eq.\eqref{g37} that the positive contribution of
$f(\mathcal{R},\mathcal{T},\mathcal{Q})$ corrections lessen the
gravitational force that ultimately diminishes the collapse rate.
However, the collapse rate may increase for negative effects of
modified terms, but its chances are much less as compared to the
former case due to the engagement of charge.

\item We have considered the matter Lagrangian in terms of the Maxwell field tensor
leading to $\mathbb{L}_{\mathfrak{M}}=\frac{s^2}{2\mathcal{Y}^4}$.
Due to the inclusion of such form of Lagrangian, the modified field
equations contain multiple charge terms along with geometric source
terms. As a result, the factor $\mathfrak{H}$ in Eq.\eqref{g37}
involves charge unlike from $\mathbb{GR}$ \cite{35c},
$f(\mathcal{R})$ \cite{h} and $f(\mathcal{R},\mathcal{T})$ \cite{i}
theories, that helps to reduce the rate of collapse.

\item The active gravitational mass also contains the effects of charge which is
inconsistent with the result found in \cite{i}.
\end{itemize}

We have further discussed the criterion for the spacetime to be
conformally flat. For this, we have linked the Weyl scalar and state
variables such as the energy density and principal pressures through
an appropriate relationship for a standard model \eqref{g5d}. By
imposing some restrictions on the model along with the regular axis
condition, we have found that the conformal flatness and the energy
density homogeneity of the considered matter source imply each
other. It must be mentioned that the existence of the tidal forces
make the fluid more inhomogeneous throughout the evolution. We
retrieve all of our findings in $\mathbb{GR}$ by substituting
$\Phi=0=\Psi$.

\section*{Appendix A}

The $f(\mathcal{R},\mathcal{T},\mathcal{Q})$ corrections in the
field equations \eqref{g8}-\eqref{g8d} are
\begin{align}\nonumber
\mu^{(D)}&=-\frac{\mathcal{P}^2\big(\Phi\sqrt{\mathcal{T}}+\Psi\mathcal{Q}\big)}{2}+\Psi\bigg\{\mu\bigg(\frac{3\ddot{\mathcal{W}}}{2\mathcal{W}^2}
-\frac{3\dot{\mathcal{P}}\dot{\mathcal{W}}}{2\mathcal{PW}}+\frac{3\ddot{\mathcal{Y}}}{\mathcal{Y}}-\frac{\dot{\mathcal{Y}}^2}{\mathcal{Y}^2}
-\frac{\dot{\mathcal{P}}\dot{\mathcal{Y}}}{\mathcal{PY}}+\frac{2\mathcal{P}'^2}{\mathcal{W}^2}\\\nonumber
&-\frac{3\mathcal{P}\mathcal{P}''}{2\mathcal{W}^2}-\frac{1}{2}\mathcal{P}^2\mathcal{R}-\frac{\mathcal{P}\mathcal{P}'\mathcal{W}'}{2\mathcal{W}^3}
-\frac{2\dot{\mathcal{W}}\dot{\mathcal{Y}}}{\mathcal{WY}}-\frac{3\mathcal{P}\mathcal{P}'\mathcal{Y}'}{\mathcal{W}^2\mathcal{Y}}\bigg)
-\dot{\mu}\bigg(\frac{3\dot{\mathcal{P}}}{\mathcal{P}}+\frac{\dot{\mathcal{Y}}}{\mathcal{Y}}
+\frac{\dot{\mathcal{W}}}{2\mathcal{W}}\bigg)\\\nonumber
&-\mu'\bigg(\frac{\mathcal{P}^2\mathcal{W}'}{2\mathcal{W}^3}-\frac{\mathcal{P}^2\mathcal{Y}'}{\mathcal{W}^2}\bigg)
+\frac{\mu''\mathcal{P}^2}{2\mathcal{W}^2}+\mathrm{P}_{r}\bigg(\frac{\mathcal{P}\mathcal{P}''}{2\mathcal{W}^2}-\frac{\ddot{\mathcal{W}}}
{2\mathcal{W}}-\frac{\mathcal{P}\mathcal{P}'\mathcal{W}'}{2\mathcal{W}^3}-\frac{\mathcal{P}^2\mathcal{Y}''}{\mathcal{W}^2\mathcal{Y}}\\\nonumber
&-\frac{\mathcal{P}^2\mathcal{Y}'^2}{\mathcal{W}^2\mathcal{Y}^2}+\frac{2\mathcal{P}^2\mathcal{W}'\mathcal{Y}'}{\mathcal{W}^3\mathcal{Y}}
-\frac{\dot{\mathcal{W}}\dot{\mathcal{Y}}}{\mathcal{WY}}-\frac{2\mathcal{P}^2\mathcal{W}'\dot{\mathcal{Y}}}{\mathcal{W}^3\mathcal{Y}}
+\frac{2\mathcal{P}'^2}{\mathcal{W}^2}\bigg)+\frac{\dot{\mathrm{P}}_{r}\dot{\mathcal{W}}}{2\mathcal{W}}
-\frac{\mathrm{P}''_{r}\mathcal{P}}{2\mathcal{W}^2}\\\nonumber
&+\mathrm{P}'_{r}\bigg(\frac{\mathcal{P}^2\mathcal{W}'}{2\mathcal{W}^3}-\frac{2\mathcal{P}^2\mathcal{Y}'}{\mathcal{W}^2\mathcal{Y}}\bigg)
-\mathrm{P}_{\bot}\bigg(\frac{\ddot{\mathcal{Y}}}{\mathcal{Y}}-\frac{\mathcal{P}\mathcal{P}'\mathcal{Y}'}{\mathcal{W}^2\mathcal{Y}}
-\frac{\dot{\mathcal{P}}\dot{\mathcal{Y}}}{\mathcal{PY}}+\frac{\mathcal{P}^2\mathcal{W}'\mathcal{Y}'}{\mathcal{W}^3\mathcal{Y}}
-\frac{\mathcal{P}^2\mathcal{Y}'^2}{\mathcal{W}^2\mathcal{Y}^2}\\\nonumber
&+\frac{\dot{\mathcal{Y}}^2}{\mathcal{Y}}-\frac{\mathcal{P}^2\mathcal{Y}''}{\mathcal{W}^2\mathcal{Y}}
+\frac{\dot{\mathcal{W}}\dot{\mathcal{Y}}}{\mathcal{WY}}\bigg)-\frac{3\dot{\mathrm{P}}_{\bot}\dot{\mathcal{Y}}}{\mathcal{Y}}
+\frac{\mathrm{P}'_{\bot}\mathcal{P}^2\mathcal{Y}'}{\mathcal{W}^2\mathcal{Y}}-\varsigma\bigg(\frac{9\dot{\mathcal{P}}\mathcal{P}'}{2\mathcal{PW}}
-\frac{2\mathcal{P}\dot{\mathcal{Y}}'}{\mathcal{WY}}+\frac{\dot{\mathcal{P}}\mathcal{Y}'}{\mathcal{WY}}\\\nonumber
&+\frac{3\mathcal{P}'\dot{\mathcal{Y}}}{\mathcal{WY}}+\frac{5\mathcal{P}\dot{\mathcal{W}}\mathcal{Y}'}{\mathcal{W}^2\mathcal{Y}}
+\frac{3\mathcal{P}'\dot{\mathcal{W}}}{\mathcal{W}^2}+\frac{2\mathcal{P}\dot{\mathcal{Y}}\mathcal{Y}'}{\mathcal{WY}^2}\bigg)
+\frac{2\dot{\varsigma}\mathcal{P}'}{\mathcal{W}}-\frac{2\varsigma'\mathcal{P}\dot{\mathcal{Y}}}{\mathcal{WY}}\bigg\},\\\nonumber
\mathrm{P}_{\mathrm{r}}^{(D)}&=\frac{\mathcal{W}^2}{2}\bigg\{\bigg(\frac{\Phi}{\sqrt{\mathcal{T}}}+\Psi\mathcal{R}\bigg)\mathrm{P}_{\mathrm{r}}
+\Phi\sqrt{\mathcal{T}}+\Psi\mathcal{Q}+\frac{\Phi\mu}{\sqrt{\mathcal{T}}}\bigg\}
+\Psi\bigg\{\mu\bigg(\frac{\mathcal{W}\ddot{\mathcal{W}}}{2\mathcal{P}^2}
-\frac{\mathcal{W}^2\ddot{\mathcal{Y}}}{\mathcal{P}^2\mathcal{Y}}\\\nonumber
&-\frac{\mathcal{W}\dot{\mathcal{P}}\dot{\mathcal{W}}}{2\mathcal{P}^3}
+\frac{\mathcal{W}^2\dot{\mathcal{P}}\dot{\mathcal{Y}}}{\mathcal{P}^3\mathcal{Y}}-\frac{\mathcal{P}''}{2\mathcal{P}}
-\frac{\mathcal{W}^2\dot{\mathcal{Y}}^2}{\mathcal{P}^2\mathcal{Y}^2}+\frac{\mathcal{P}'\mathcal{W}'}{2\mathcal{P}\mathcal{W}}
-\frac{\mathcal{P}'\mathcal{Y}'}{\mathcal{P}\mathcal{Y}}\bigg)+\frac{\mu'\mathcal{P}'}{2\mathcal{P}}
-\frac{\ddot{\mu}\mathcal{W}^2}{2\mathcal{P}^2}\\\nonumber
&+\dot{\mu}\bigg(\frac{\mathcal{W}^2\dot{\mathcal{P}}}{2\mathcal{P}^3}-\frac{2\mathcal{W}^2\dot{\mathcal{Y}}}{\mathcal{P}^2\mathcal{Y}}\bigg)
+\mathrm{P}_{r}\bigg(\frac{1}{2}\mathcal{W}^2\mathcal{R}+\frac{3\mathcal{W}\dot{\mathcal{P}}\dot{\mathcal{W}}}{2\mathcal{P}^3}
-\frac{3\mathcal{W}\ddot{\mathcal{W}}}{2\mathcal{P}^2}-\frac{2\mathcal{P}'\mathcal{Y}'}{\mathcal{P}\mathcal{Y}}
+\frac{3\mathcal{P}''}{2\mathcal{P}}\\\nonumber
&-\frac{3\mathcal{P}'\mathcal{W}'}{2\mathcal{P}\mathcal{W}}-\frac{2\mathcal{W}'\mathcal{Y}'}{\mathcal{W}\mathcal{Y}}
+\frac{3\mathcal{Y}''}{\mathcal{Y}}-\frac{3\mathcal{W}\dot{\mathcal{W}}\dot{\mathcal{Y}}}{\mathcal{P}^2\mathcal{Y}}
-\frac{2\mathcal{W}'\dot{\mathcal{Y}}}{\mathcal{W}\mathcal{Y}}-\frac{\mathcal{Y}'^2}{\mathcal{Y}^2}\bigg)
-\mathrm{P}'_{r}\bigg(\frac{\mathcal{P}'}{2\mathcal{P}}+\frac{\mathcal{Y}'}{\mathcal{Y}}\bigg)\\\nonumber
&+\dot{\mathrm{P}}_{r}\bigg(\frac{\mathcal{W}^2\dot{\mathcal{Y}}}{\mathcal{P}^2\mathcal{Y}}-\frac{\mathcal{W}^2\dot{\mathcal{P}}}{2\mathcal{P}^3}
\bigg)+\frac{\ddot{\mathrm{P}}_{r}\mathcal{W}^2}{2\mathcal{P}^2}+\mathrm{P}_{\bot}\bigg(\frac{\mathcal{P}'\mathcal{Y}'}{\mathcal{P}\mathcal{Y}}
-\frac{\mathcal{W}^2\dot{\mathcal{Y}}^2}{\mathcal{P}^2\mathcal{Y}^2}
+\frac{\mathcal{W}^2\dot{\mathcal{P}}\dot{\mathcal{Y}}}{\mathcal{P}^3\mathcal{Y}}+\frac{\mathcal{Y}''}{\mathcal{Y}}\\\nonumber
&-\frac{\mathcal{W}^2\ddot{\mathcal{Y}}}{\mathcal{P}^2\mathcal{Y}}-\frac{\mathcal{W}'\mathcal{Y}'}{\mathcal{W}\mathcal{Y}}
+\frac{\mathcal{Y}'^2}{\mathcal{Y}^2}-\frac{\mathcal{W}\dot{\mathcal{W}}\dot{\mathcal{Y}}}{\mathcal{P}^2\mathcal{Y}}\bigg)
-\frac{\dot{\mathrm{P}}_{\bot}\mathcal{W}^2\dot{\mathcal{Y}}}{\mathcal{P}^2\mathcal{Y}}-\frac{\mathrm{P}'_{\bot}\mathcal{Y}'}{\mathcal{Y}}
+\varsigma\bigg(\frac{2\mathcal{W}\dot{\mathcal{Y}}'}{\mathcal{P}\mathcal{Y}}\\\nonumber
&-\frac{\mathcal{W}\dot{\mathcal{P}}\mathcal{Y}'}{\mathcal{P}^2\mathcal{Y}}
-\frac{3\mathcal{W}\mathcal{P}'\dot{\mathcal{Y}}}{\mathcal{P}^2\mathcal{Y}}-\frac{2\mathcal{W}\dot{\mathcal{Y}}\mathcal{Y}'}{\mathcal{PY}^2}
-\frac{4\dot{\mathcal{W}}\mathcal{Y}'}{\mathcal{P}\mathcal{Y}}\bigg)-\frac{2\dot{\varsigma}\mathcal{WY}'}{\mathcal{PY}}\bigg\},\\\nonumber
\mathrm{P}_{\bot}^{(D)}&=\frac{\mathcal{Y}^2}{2}\bigg\{\bigg(\frac{\Phi}{\sqrt{\mathcal{T}}}+\Psi\mathcal{R}\bigg)\mathrm{P}_{\bot}
+\Phi\sqrt{\mathcal{T}}+\Psi\mathcal{Q}+\frac{\Phi\mu}{\sqrt{\mathcal{T}}}\bigg\}
+\Psi\bigg\{\mu\bigg(\frac{\mathcal{Y}^2\dot{\mathcal{P}}\dot{\mathcal{W}}}{2\mathcal{P}^3\mathcal{W}}\\\nonumber
&-\frac{\mathcal{Y}^2\ddot{\mathcal{W}}}{2\mathcal{P}^2\mathcal{W}}-\frac{\mathcal{Y}^2\mathcal{P}''}{2\mathcal{P}\mathcal{W}^2}
+\frac{\mathcal{Y}^2\mathcal{P}'\mathcal{W}'}{2\mathcal{P}\mathcal{W}^3}
-\frac{\mathcal{Y}\dot{\mathcal{W}}\dot{\mathcal{Y}}}{\mathcal{P}^2\mathcal{W}}
-\frac{\mathcal{Y}\mathcal{P}'\mathcal{Y}'}{\mathcal{P}\mathcal{W}^2}\bigg)
-\frac{\mu'\mathcal{Y}^2\mathcal{P}'}{2\mathcal{P}\mathcal{W}^2}-\frac{\ddot{\mu}\mathcal{Y}^2}{2\mathcal{P}^2}\\\nonumber
&+\dot{\mu}\bigg(\frac{\mathcal{Y}^2\dot{\mathcal{P}}}{2\mathcal{P}^3}-\frac{\mathcal{Y}^2\dot{\mathcal{W}}}{\mathcal{P}^2\mathcal{W}}
-\frac{\mathcal{Y}\dot{\mathcal{Y}}}{\mathcal{P}^2}\bigg)
+\mathrm{P}_{r}\bigg(\frac{\mathcal{Y}^2\dot{\mathcal{P}}\dot{\mathcal{W}}}{2\mathcal{P}^3\mathcal{W}}
-\frac{\mathcal{Y}^2\ddot{\mathcal{W}}}{2\mathcal{P}^2\mathcal{W}}-\frac{\mathcal{Y}\mathcal{P}'\mathcal{Y}'}{\mathcal{P}\mathcal{W}^2}
-\frac{\mathcal{Y}^2\mathcal{P}''}{2\mathcal{P}\mathcal{W}^2}\\\nonumber
&+\frac{\mathcal{Y}^2\mathcal{P}'\mathcal{W}'}{2\mathcal{P}\mathcal{W}^3}+\frac{\mathcal{Y}\mathcal{W}'\mathcal{Y}'}{\mathcal{W}^3}
-\frac{\mathcal{Y}\dot{\mathcal{W}}\dot{\mathcal{Y}}}{\mathcal{P}^2\mathcal{W}}
-\frac{2\mathcal{Y}\mathcal{W}'\dot{\mathcal{Y}}}{\mathcal{W}^3}\bigg)-\frac{\dot{\mathrm{P}}_{r}\mathcal{Y}^2\dot{\mathcal{W}}}
{2\mathcal{P}^2\mathcal{W}}-\frac{\mathrm{P}''_{r}\mathcal{Y}^2}{2\mathcal{W}^2}\\\nonumber
&+\mathrm{P}'_{r}\bigg(\frac{\mathcal{Y}^2\mathcal{W}'}{2\mathcal{W}^3}-\frac{\mathcal{Y}^2\mathcal{P}'}{\mathcal{P}\mathcal{W}^2}
-\frac{\mathcal{Y}\mathcal{Y}'}{\mathcal{W}^2}\bigg)+\mathrm{P}_{\bot}\bigg(\frac{1}{2}\mathcal{Y}^2\mathcal{R}
+\frac{2\mathcal{Y}\mathcal{Y}''}{\mathcal{W}^2}-\frac{2\mathcal{Y}\ddot{\mathcal{Y}}}{\mathcal{P}^2}
-\frac{2\dot{\mathcal{Y}}^2}{\mathcal{P}^2}\\\nonumber
&+\frac{2\mathcal{Y}'^2}{\mathcal{W}^2}-\frac{2\mathcal{Y}\mathcal{W}'\mathcal{Y}'}{\mathcal{W}^3}
+\frac{2\mathcal{Y}\mathcal{P}'\mathcal{Y}'}{\mathcal{P}\mathcal{W}^2}
-\frac{2\mathcal{Y}\dot{\mathcal{W}}\dot{\mathcal{Y}}}{\mathcal{P}^2\mathcal{W}}-2
+\frac{2\mathcal{Y}\dot{\mathcal{P}}\dot{\mathcal{Y}}}{\mathcal{P}^3}\bigg)+\dot{\mathrm{P}}_{\bot}\mathcal{Y}^2\\\nonumber
&\times\bigg(\frac{\dot{\mathcal{W}}}{2\mathcal{P}^2\mathcal{W}}-\frac{\dot{\mathcal{P}}}{2\mathcal{P}^3}\bigg)
+\mathrm{P}'_{\bot}\mathcal{Y}^2\bigg(\frac{\mathcal{W}'}{2\mathcal{W}^3}-\frac{\mathcal{P}'}{2\mathcal{P}\mathcal{W}^2}\bigg)
+\frac{\ddot{\mathrm{P}}_{\bot}\mathcal{Y}^2}{2\mathcal{P}^2}-\frac{\mathrm{P}''_{\bot}\mathcal{Y}^2}{2\mathcal{W}^2}\\\nonumber
&+\varsigma\bigg(\frac{\mathcal{Y}^2\dot{\mathcal{P}}\mathcal{P}'}{\mathcal{P}^3\mathcal{W}}
-\frac{\mathcal{Y}^2\dot{\mathcal{P}}'}{\mathcal{P}^2\mathcal{W}}-\frac{\mathcal{Y}\dot{\mathcal{P}}\mathcal{Y}'}{\mathcal{P}^2\mathcal{W}}
+\frac{\mathcal{Y}^2\dot{\mathcal{W}}\mathcal{W}'}{\mathcal{P}\mathcal{W}^3}-\frac{\mathcal{Y}^2\dot{\mathcal{W}}'}{\mathcal{P}\mathcal{W}^2}
-\frac{\mathcal{Y}\mathcal{P}'\dot{\mathcal{Y}}}{\mathcal{P}^2\mathcal{W}}\\\nonumber
&-\frac{2\mathcal{Y}\dot{\mathcal{W}}\mathcal{Y}'}{\mathcal{P}\mathcal{W}^2}\bigg)
-\dot{\varsigma}\bigg(\frac{\mathcal{Y}^2\mathcal{P}'}{\mathcal{P}^2\mathcal{W}}+\frac{\mathcal{Y}\mathcal{Y}'}{\mathcal{P}\mathcal{W}}\bigg)
-\varsigma'\bigg(\frac{\mathcal{Y}^2\dot{\mathcal{W}}}{\mathcal{P}\mathcal{W}^2}+\frac{\mathcal{Y}\dot{\mathcal{Y}}}{\mathcal{P}\mathcal{W}}\bigg)
-\frac{\dot{\varsigma}'\mathcal{Y}^2}{\mathcal{P}\mathcal{W}}\bigg\},\\\nonumber
\varsigma^{(D)}&=-\frac{\varsigma\mathcal{PW}}{2}\bigg(\frac{\Phi}{\sqrt{\mathcal{T}}}+\Psi\mathcal{R}\bigg)
+\Psi\bigg\{\mu\bigg(\frac{2\dot{\mathcal{Y}}'}{\mathcal{Y}}-\frac{\dot{\mathcal{P}}\mathcal{P}'}{\mathcal{P}^2}
-\frac{3\dot{\mathcal{P}}'}{2\mathcal{P}}-\frac{2\dot{\mathcal{W}}\mathcal{Y}'}{\mathcal{W}\mathcal{Y}}
-\frac{2\mathcal{P}'\dot{\mathcal{Y}}}{\mathcal{P}\mathcal{Y}}\bigg)\\\nonumber
&-\frac{3\dot{\mu}\mathcal{P}'}{2\mathcal{P}}+\mu'\bigg(\frac{\dot{\mathcal{Y}}}{\mathcal{Y}}-\frac{\dot{\mathcal{W}}}{2\mathcal{W}}\bigg)
+\frac{\dot{\mu}'}{2}+\mathrm{P}_{r}\bigg(\frac{\mathcal{P}'^2}{2\mathcal{W}^2}-\frac{\mathcal{P}'\dot{\mathcal{W}}}{2\mathcal{P}\mathcal{W}}
+\frac{2\dot{\mathcal{W}}\mathcal{Y}'}{\mathcal{W}\mathcal{Y}}+\frac{2\mathcal{P}'\dot{\mathcal{Y}}}{\mathcal{P}\mathcal{Y}}\\\nonumber
&-\frac{2\dot{\mathcal{Y}}'}{\mathcal{Y}}\bigg)+\dot{\mathrm{P}}_{r}\bigg(\frac{\mathcal{P}'}{2\mathcal{P}}-\frac{\mathcal{Y}'}{\mathcal{Y}}\bigg)
+\frac{\mathrm{P}'_{r}\dot{\mathcal{W}}}{2\mathcal{W}}-\frac{\dot{\mathrm{P}}'_{r}}{2}+\frac{\dot{\mathrm{P}}_{\bot}\mathcal{Y}'}{\mathcal{Y}}
+\frac{\mathrm{P}'_{\bot}\dot{\mathcal{Y}}}{\mathcal{Y}}+\varsigma\bigg(\frac{1}{2}\mathcal{P}\mathcal{W}\mathcal{R}\\\nonumber
&-\frac{2\mathcal{W}\ddot{\mathcal{Y}}}{\mathcal{P}\mathcal{Y}}-\frac{\ddot{\mathcal{W}}}{\mathcal{P}}+\frac{\mathcal{P}''}{\mathcal{W}}
+\frac{2\mathcal{W}\dot{\mathcal{P}}\dot{\mathcal{Y}}}{\mathcal{P}^2\mathcal{Y}}+\frac{2\mathcal{P}\mathcal{Y}''}{\mathcal{W}\mathcal{Y}}
-\frac{2\mathcal{P}\mathcal{W}'\mathcal{Y}'}{\mathcal{W}^2\mathcal{Y}}+\frac{\dot{\mathcal{P}}\dot{\mathcal{W}}}{\mathcal{P}^2}
-\frac{\mathcal{P}'\mathcal{W}'}{\mathcal{W}^2}\bigg)\bigg\},\\\nonumber
\mathcal{E}_{00}^{(D)}&=\frac{\Psi{s}^2}{\mathcal{PW}^3\mathcal{Y}^5}\left(2\mathcal{P}^3\mathcal{WY}''
-\mathcal{PW}^2\mathcal{Y}\ddot{\mathcal{W}}+\mathcal{P}^2\mathcal{WYP}''-2\mathcal{P}^3\mathcal{W}'\mathcal{Y}'
-2\mathcal{PW}^2\dot{\mathcal{W}}\dot{\mathcal{Y}}\right.\\\nonumber
&+\left.\mathcal{W}^2\mathcal{Y}\dot{\mathcal{P}}\dot{\mathcal{W}}-\mathcal{P}^2\mathcal{YP}'\mathcal{W}'\right)
-\frac{\Phi{s}^2\mathcal{P}^2}{2\sqrt{\mathcal{T}}\mathcal{Y}^4},\\\nonumber
\mathcal{E}_{11}^{(D)}&=\frac{\Psi{s}^2}{\mathcal{P}^3\mathcal{WY}^5}\left(2\mathcal{PW}^3\ddot{\mathcal{Y}}+\mathcal{PW}^2\mathcal{Y}
\ddot{\mathcal{W}}-\mathcal{P}^2\mathcal{WYP}''-2\mathcal{P}^2\mathcal{WP}'\mathcal{Y}'
-2\mathcal{W}^3\dot{\mathcal{P}}\dot{\mathcal{Y}}\right.\\\nonumber
&-\left.\mathcal{W}^2\mathcal{Y}\dot{\mathcal{P}}\dot{\mathcal{W}}+\mathcal{P}^2\mathcal{YP}'\mathcal{W}'\right)
+\frac{\Phi{s}^2\mathcal{W}^2}{2\sqrt{\mathcal{T}}\mathcal{Y}^4},\\\nonumber
\mathcal{E}_{01}^{(D)}&=\frac{\Psi{s}^2\mathcal{PW}}{\mathcal{Y}^5}\left(\mathcal{P}\dot{\mathcal{W}}\mathcal{Y}'
+\mathcal{W}\mathcal{P}'\dot{\mathcal{Y}}-\mathcal{PW}\dot{\mathcal{Y}}'\right).
\end{align}

\section*{Appendix B}

The Ricci tensor and the Riemann tensor for geometry \eqref{g6} are
\begin{align}\nonumber
\mathcal{R}_{00}&=\mathcal{P}^2\bigg(\frac{\mathcal{G}_{00}}{2\mathcal{P}^2}+\frac{\mathcal{G}_{11}}{2\mathcal{W}^2}
+\frac{\mathcal{G}_{22}}{\mathcal{Y}^2}\bigg),\\\nonumber
\mathcal{R}_{11}&=\mathcal{W}^2\bigg(\frac{\mathcal{G}_{00}}{2\mathcal{P}^2}+\frac{\mathcal{G}_{11}}{2\mathcal{W}^2}
-\frac{\mathcal{G}_{22}}{\mathcal{Y}^2}\bigg),\\\nonumber
\mathcal{R}_{22}&=\frac{\mathcal{R}_{33}}{sin^2\theta}=\mathcal{Y}^2\bigg(\frac{\mathcal{G}_{00}}{2\mathcal{P}^2}
-\frac{\mathcal{G}_{11}}{2\mathcal{W}^2}\bigg), \quad
\mathcal{R}_{01}=\mathcal{G}_{01},\\\nonumber
\mathcal{R}_{0101}&=\mathcal{P}^2\mathcal{W}^2\bigg\{\frac{\mathcal{G}_{00}}{2\mathcal{P}^2}-\frac{\mathcal{G}_{11}}{2\mathcal{W}^2}
+\frac{\mathcal{G}_{22}}{\mathcal{Y}^2}-\frac{2}{\mathcal{Y}^3}\bigg(\tilde{m}-\frac{s^2}{2\mathcal{Y}}\bigg)\bigg\},\\\nonumber
\mathcal{R}_{0202}&=\frac{\mathcal{R}_{0303}}{sin^2\theta}=\mathcal{P}^2\mathcal{Y}^2\bigg\{\frac{\mathcal{G}_{11}}{2\mathcal{W}^2}
+\frac{1}{\mathcal{Y}^3}\bigg(\tilde{m}-\frac{s^2}{2\mathcal{Y}}\bigg)\bigg\},\\\nonumber
\mathcal{R}_{1212}&=\frac{\mathcal{R}_{1313}}{sin^2\theta}=\mathcal{W}^2\mathcal{Y}^2\bigg\{\frac{\mathcal{G}_{00}}{2\mathcal{P}^2}
-\frac{1}{\mathcal{Y}^3}\bigg(\tilde{m}-\frac{s^2}{2\mathcal{Y}}\bigg)\bigg\},\\\nonumber
\mathcal{R}_{2323}&=2\mathcal{Y}\bigg(\tilde{m}-\frac{s^2}{2\mathcal{Y}}\bigg)sin^2\theta,\quad
\mathcal{R}_{0212}=\frac{\mathcal{R}_{0313}}{sin^2\theta}=\frac{\mathcal{Y}^2\mathcal{G}_{01}}{2}.
\end{align}

\end{document}